\documentclass[aps,pra,preprint,groupedaddress]{revtex4-2}
\usepackage{amsmath}
\usepackage{amssymb}
\usepackage[pdftex]{graphicx}

\usepackage{tabularx}
\usepackage{footmisc}

\begin{document}

%\preprint{}

\title{Theory of Laser Induced Strong-Field Ionization Tomographic Imaging}


\author{Eugene Frumker}
\email[]{efrumker@bgu.ac.il}
%\thanks{}
\affiliation{Department of Physics, Ben-Gurion University of the Negev, Beer-Sheva 84105, Israel}


%\date{\today}

\begin{abstract}

Laser-induced strong-field ionization tomography (LISFIT) is an emerging space-time tomographic modality with the potential to revolutionize imaging capabilities.  To fully harness its power, a robust theoretical framework is essential. This work delves into the fundamental physics of strong-field ionization and its implications for tomographic imaging.

Our analysis reveals an operational regime with significant resolution enhancement and unique intensity-resolution coupling, alongside  localization phenomena rooted in the physics of strong-field interactions. We further identify a trade-off between resolution, localization extent, and signal-to-noise ratio, providing critical insights for optimizing experimental parameters.
\end{abstract}

%\keywords{}
%\keywords{Resolution, Strong field laser physics, Tomography, Ultrafast nonlinear optics}

\maketitle

\section{\label{Intro} Introduction}

In recent decades, advancements in ultrafast laser science and technology have revolutionized imaging and microscopy. A new paradigm centered around non-linear laser scanning techniques has emerged, offering significant advantages over traditional linear methods.

These imaging techniques harness various perturbative nonlinear interactions driven by focused laser beams, including second harmonic generation (SHG) \cite{Gannaway_Scanning_Micr_SHG_OQE_1978}, two-photon fluorescence (TPF) \cite{Denk_TPF_microscopy_Science_1990}, third harmonic generation (THG) \cite{Barad_THG_microscopy_APL1997}, and stimulated Raman scattering (SRS) \cite{Freudiger_SRS_Microscopy_Science_2008}, among others. By exploiting these non-linear processes, researchers have achieved improved resolution and optical sectioning capabilities, enabling deeper insights into biological and material systems.

Another pivotal advancement that extended imaging from two to three dimensions is computed tomography (CT) \cite{Kak_book_tomography_2002}. Tomographic imaging reconstructs an image from multiple projections, enabling to capture of detailed 3D structures. This approach has had a transformative impact across diverse fields, including physics, materials science, life sciences, atmospheric science, and medicine.

Recently, we introduced non-perturbative 3D imaging approach: laser-induced strong-field ionization tomography (LISFIT) \cite{Tchulov_LaserStrongFieldTom_ScRep_2017}. This method was further extended to a dynamic four-dimensional (4D = 3D space +1D time) appproach \cite{Shlomo_4D_tomography_arXiv_2024}. 
To the best of our knowledge, this represents the first application of strong-field nonlinear interactions for imaging. 
In its current form, LISFIT has been successfully applied to image supersonic gas jets.
 However, its potential applications extend far beyond this initial demonstration.

To facilitate the widespread adoption and maximize the potential of LISFIT,  establishing a comprehensive  theoretical framework for these new imaging modalities is essential. In this work, we present the theoretical foundations of laser-induced strong-field ionization imaging and tomography, discussing its key characteristics and potential application benefits.

\section{Strong-field ionization in the laser field}

\subsection{Key Theoretical Milestones in Strong-Field Ionization}

\label{Section_Key_milestones}

 The ionization of atoms and molecules subject to a strong laser field has been extensively studied  both theoretically and experimentally for over half a century \cite{Keldysh_ionization_JETP1965, Perelomov_PPT_JETP1966, Ammosov_ADK_JETP_1986, Eckle_tunneling_time_delay_Science_2008, Sainadh_Litvinyuk_tunnelling_time_in_H_Nature_2019}.
 %\cite{Keldysh_ionization_JETP1965, Perelomov_PPT_JETP1967, Ammosov_ADK_JETP_1986, Eckle_tunneling_time_delay_Science_2008, Sainadh_Litvinyuk_tunnelling_time_in_H_Nature_2019}.
In this section we will briefly recount key milestones of the strong field ionization theory with an emphasis on the  aspects pertaining to our further analysis.

The theoretical foundation for strong-field ionization was laid by Keldysh \cite{Keldysh_ionization_JETP1965}, who explored nonlinear ionization in the "transparency region," where the ionization potential $I_p$ exceeds the photon energy $\hbar \omega_L$. Keldysh introduced the adiabaticity, or Keldysh parameter ($\gamma = 2 \omega_L T_{\textrm{tunnel}} = 4 \pi \frac{T_{\textrm{tunnel}}}{T_L}$), which defines the nature of the process as the ratio of two key timescales: the laser period $T_L = 2 \pi / \omega_L$ and the so-called "tunneling time" $T_{\textrm{tunnel}}$.
%The tunneling time is defined as $T_{\textrm{tunnel}}=\textrm{l}_\textrm{barrier} /  \textrm{v}_{\textrm{e}} = \sqrt{I_p m_e}/ {\sqrt{2} e E_L}$,
% where $\textrm{l}_\textrm{barrier}= I_p /{e E_L}$ - is the length of potential barrier (replacing the Coulomb with point-like short range potential). To estimate the average velocity of the electron under the barrier, we note that  using virial theorem ( $E_K = m \textrm{v}_{\textrm{e}}^2/2 = I_p$) for Coulomb-like ($\sim 1/r$) potential, hence $\textrm{v}_{\textrm{e}^{max}} = \sqrt{2 I_p / m_e}$.

To define the tunneling time \cite{Keldysh_ionization_JETP1965}, we first estimate the "classical" velocity of electron orbiting the nuclear under the barrier, using virial theorem \cite{Clausius_Virial_Theorem_PhilMag_1870}: 
for a Coulomb-like potential ($\sim 1/r$), the electron’s kinetic energy $E_K$ is related to its velocity $\textrm{v}_{\textrm{e}}$ as $E_K = m_{\textrm{e}} \textrm{v}_{\textrm{e}}^2/2 = I_p$, where $m_{\textrm{e}}$ is the electron mass. This gives $\textrm{v}_{\textrm{e}} = \sqrt{2I_p / m_{\textrm{e}}}$
\footnote{\label{note1_Keldysh}Note, that in the original Keldysh paper \cite{Keldysh_ionization_JETP1965}, only an estimate was given for the average electron velocity as  $(I_p/m)^{1/2}$ that differs by a factor of $\sqrt{2}$ from the exact value. Hence there is a noncritical discrepancy (within a factor of 2) appears in definitions of the tunneling time, the tunneling barrier length and the Keldysh parameter  between different sources.}.
  
\begin{figure}[hbpt]
    \begin{center}
\includegraphics[width=\linewidth]{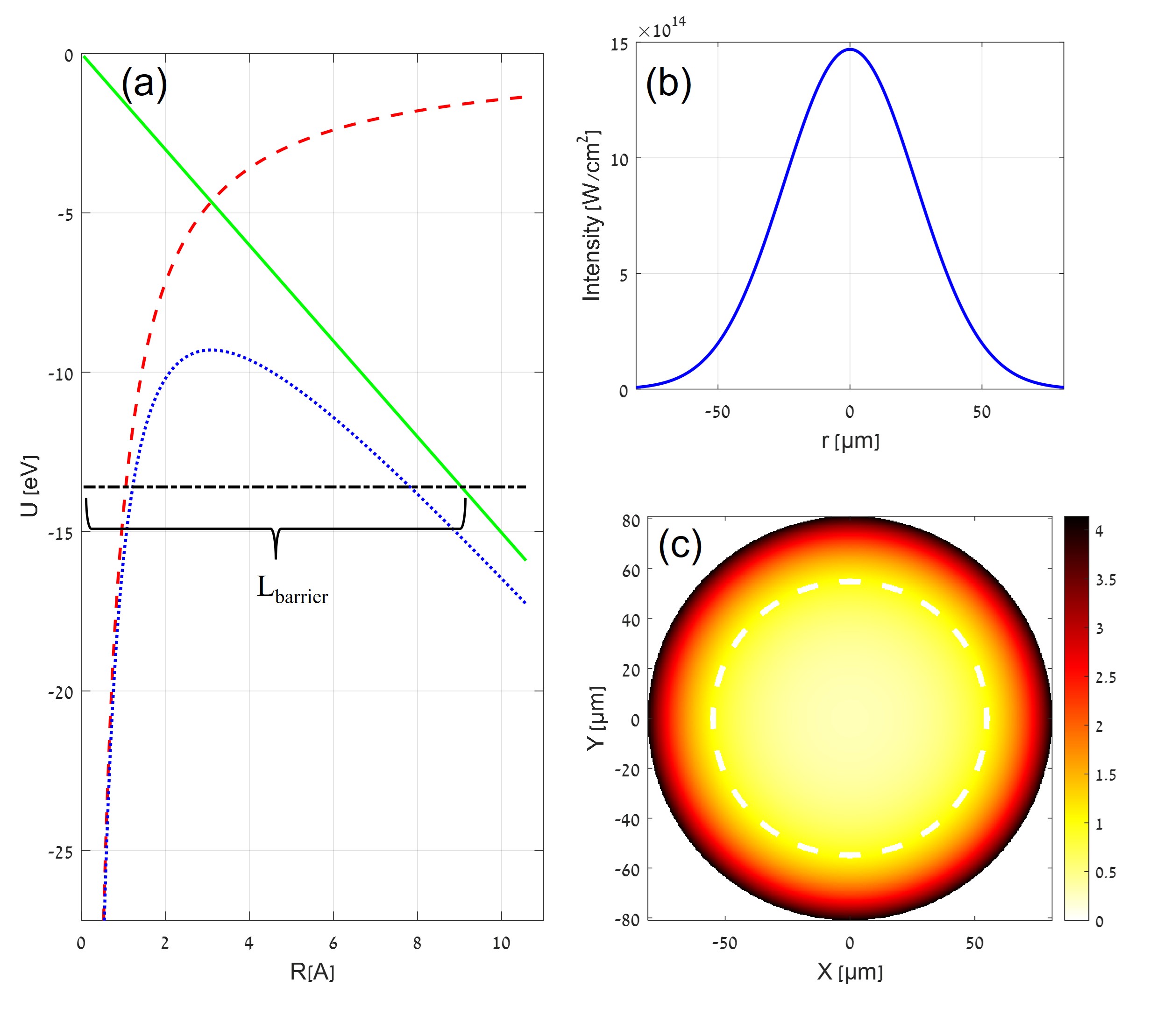}
%G:\BGU_since_2013\papers\Gas_Jet_ionization_Tomography\Numerical_evaluations_Gas_Jet_ionization\Figures_making4Theory_Gas_Jet_ionization_paper.m

    %\captionsetup{justification=justified}

    \caption{ \label{fig_ionization} (a) Potential energy of electron in the hydrogen atom exposed to the laser field: dashed red curve depicts the Coulomb potential, solid green line shows the linear potential due to instantaneous laser electric field and the dotted blue curve illustrates the total potential energy of the electron. Dotted-dashed black horizontal line corresponds to the ionization potential level of hydrogen $I_p=13.6$ ev. (b) Typical intensity distribution of the focused Gaussian laser beam with a waist $w_0=50 \mu m$ . (c) Keldysh parameter map, $\gamma(X, Y)$, within the focused intense Gaussian beam as shown in (b). The dashed white circle shows the $\gamma(X, Y)=1$ curve that separate between the tunnel-ionization regime in the vicinity of the peak of the beam and the multiphoton ionization regime occurring in the periphery of the beam.}

    \end{center}
\end{figure}

  In this picture, we may think of the electron tunneling through a potential barrier formed by the superposition of the laser electric field  $E_L$ potential and the Coulomb attraction potential as schematically shown in Fig. \ref{fig_ionization} (a). 
  To further simplify the problem, it is common to assume a short range, point-like potential of an electron interaction with the nuclear and approximate the total potential by a triangular barrier shown in Fig. \ref{fig_ionization} (a) by the solid  line. 
  For an electron to escape, it must traverse a distance $\textrm{L}_\textrm{barrier} = I_p / (e E_L)$, where $e$ is the electron charge. 
  The time it takes for an electron moving with the velocity $\textrm{v}_{\textrm{e}}$ to traverse the potential barrier distance $\textrm{L}_\textrm{barrier}$ is defined as the tunnelling time:
$T_{\textrm{tunnel}}=\frac{\textrm{L}_\textrm{barrier}}{{\textrm{v}}_{\textrm{e}}}=\frac{\sqrt{I_p m_{\textrm{e}}}}{\sqrt{2} e E_L}$.
%[??ref Milonni Sundaram "Progress in Optics" Atoms in Strong field: Photoionization and chaos, %\verbatim{Popruzhenko_Keldysh_theory_strong_field_ionization__J._Phys._B_2014},??]

%For the progress in Optics reference please see email from Oshrat:
%https://mail.google.com/mail/u/1/#inbox/WhctKJVjSkDMQCJHXprzvsQtvcgddVTnFNbzthsKGqbdGVbnVqhbzLsrHCbrzkWnCgFPtjG

%Assuming that electron emerges into the continua with zero velocity ([??cite Ivanov Ivanov "Anatomy of strong field ionization" 2005??], we have $\textrm{v}_{\textrm{e}}-(e E_L /m_e)*T_{\textrm{tunnel}}=0$, hence $T_{\textrm{tunnel}}=\frac{\sqrt{2I_p m_e}}{ e E_L}$.
%To complete this naive picture, we may think of the electron traveling through constant laser field created by triangular barrier and Coulomb potential (approximated by short range point-like potential ) of the length $\textrm{l}_\textrm{barrier}= I_p /{e E_L}$ as shown in Fig. ??.
%Hence the average velocity under the barrier can be estimated $\overline{\textrm{v}}_{\textrm{e}}=\textrm{l}_\textrm{barrier}/T_{\textrm{tunnel}}=\sqrt{I_p /2 m_e}$

 %G:\NRC_work_since_December2007\papers_NRC\Tunnelling_and_ionization
 %Misha Ivanov makes very strange estimation of tunneling time in his paper (page 3) "From tunnel to multiphoton ionization: Looking inside a laser cycle": For the binding energy I_p the velocity..."

It is worth mentioning that the very notion of tunneling time, its physical significance, and attempts to measure it have been sources of ongoing debate in the community since the early days of quantum mechanics (for some examples, see \cite{Landauer_Barrier_interaction_time__tunneling_Rev_Mod_Phys_1994, Greenberger_Fundamental_problems_quantum_theory_Conf_1995, Ivanov_Anatomy_J_Mod_Opt2005, Eckle_tunneling_time_delay_Science_2008, Zheltikov_Keldysh_tunneling_time_PRA_2016, Sainadh_Litvinyuk_tunnelling_time_in_H_Nature_2019}). In this work, we will avoid the temptation to delve into this discussion and will use the term "tunneling time" strictly by its aforementioned definition and its implications within the framework of the theory of strong field ionization.
Thus, we obtain for the Keldysh parameter:
%For the definition of Keldysh parameters via tunneling time see Wiki:
%https://en.wikipedia.org/wiki/Tunnel_ionization

\begin{equation}
\label{Keldysh_parameter}
\gamma  = 2 \pi \frac{\sqrt{2I_p m_{\textrm{e}}}}{ e E_L T_L}.
\end{equation}

%[?? There is a problem of about $2\pi$ factor with "common" definition of Keldysh parameter ?? - find why. Compare with the Keldysh paper.
%Partial answer - Keldysh incorrectly estimated the velocity of electron under the barrier with ionization potential Ip??]

For a specified laser frequency, a weaker laser field $E_L$ corresponds to a longer potential barrier length $\textrm{L}_\textrm{barrier}$.
Consequently, the Keldysh parameter $\gamma$ increases. 
In the limit of $\gamma \gg 1$, or equivalently, $E_L \ll \sqrt{2I_p m_{\textrm{e}}} \omega_L/e$, the field oscillates many times during tunnelling time.
 This regime, known as the multiphoton regime, exhibits a power-law dependence of the ionization rate $W_i$  on the electric field: $W_i\propto E_L^{2K}$,
 where $K=\lfloor I_p/\omega_L+1 \rfloor$ is the minimal number of absorbed photons required by energy conservation (the  multiquantumness).
This  behavior  is reminiscent of perturbative nonlinear optical processes \cite{Boyd_NLO}.

%For a given laser frequency, a weaker laser field, $E_L$, results in a longer potential barrier length, $\textrm{L}_\textrm{barrier}$, and consequently, a higher Keldysh parameter,  $\gamma$. In the limit of $\gamma \gg 1$ or $E_L \ll \sqrt{2I_p m_{\textrm{e}}} \omega_L/e$, the field oscillates many times during tunnelling time. This is so-called multiphoton regime, where the ionization rate, $W_i$,  exhibits a power-law dependence on the electric field, familiar from perturbative nonlinear optics \cite{Boyd_NLO}: $W_i\propto E_L^{2K}$, where $K=\lfloor I_p/\omega_L+1 \rfloor$ is the minimal number of absorbed photons required by energy conservation (the  multiquantumness).

 In the opposite limiting case, when $\gamma \ll 1$ or $E_L \gg \sqrt{2I_p m_{\textrm{e}}} \omega_L /e$, the field of the driving laser doesn't change much during the tunneling time.
 In this well-known tunneling regime, the ionization rate $W_i$ for a hydrogen atom in Gaussian units  is given by \cite{Landau_QM_V3}:

%Make calculations using Landau fromula, but in SI units as appear in Wiki:
%https://en.wikipedia.org/wiki/Tunnel_ionization

 \begin{equation}
\label{Landau_ionization}
 W_i = \frac{4 m_{\textrm{e}} ^3 e^9}{E_L \hbar ^ 7} \exp[-2E_a/(3E_L)],
 \end{equation}

  where $E_a={m_{\textrm{e}}}^2 e^5/\hbar^4$ is the electrical field experienced by an electron situated at the Bohr radius $a_0=\frac{\hbar^2}{m_{\textrm{e}} e^2}$ in a hydrogen atom.

 %({?? [??Landau, paragraph 77, Problem 1, formula 5)??]; check this formula compare to Landau. May be mention that it was derived for Hydrogen atom. See also: N B Delone, V P Krainov, Tunneling and barrier-suppression ionization of atoms and ions in a laser radiation field", Physics - Uspekhi 41 (5) 469 ± 485 (1998)}  ??  )

 %In the multiphoton ($\gamma\gg 1$) regime, the dependence of the ionization rate, $W_i$, on the electric field of the optical wave is a power law: $W_i\propto E_L^{2K}$, where $K=\lfloor I_p/\omega_L+1 \rfloor$ is the threshold number of absorbed photons required by energy conservation. In the tunnel-ionization  ($\gamma\ll 1$) regime, the nonlinear ionization rate depends exponentially on $E_L$: $W_i\propto exp(-2E_a/(3E_L))$, where $E_a=(2I_p)^{3/2}$ is the characteristic atomic electric field strength. The rate formula in this case can be obtained by averaging the equation for tunneling ionization in a constant electric field over half cycle of the alternating electric field of the optical wave\cite{delone2012multiphoton}.

 Keldysh, in his work \cite{Keldysh_ionization_JETP1965}, demonstrated that both multiphoton and tunneling ionization regimes are limiting cases of the same general phenomenon of strong field ionization. This phenomenon is determined by three key parameters: the laser field strength $E_L$, the laser frequency $\omega_L$, and the ionization potential $I_p$. Keldysh proposed a unified theory for an arbitrary $\gamma$.

 Note that often in real experiment conditions,  we are neither in pure tunnelling ($\gamma \ll 1$) nor in pure multiphoton ( $\gamma \gg 1$) regimes.
 For example, in the process of HHG, the first step is ionization of the electron from the parent atom or molecule, typically occurring at the center of the driving Gaussian laser beam with Keldysh parameter on order of unity ($\gamma \simeq 1$).  This intermediate regime is particular rich and interesting as it contains signatures of both tunnelling and multiphoton ionization mechanisms.
 If we further increase the laser intensity, then across the beam, we transition from predominantly tunneling ($\gamma < 1$) through intermediate ($\gamma \simeq 1$) to multiphoton ionization ($\gamma > 1$) regimes as the Gaussian beam strength diminishes across the beam, as shown in Fig. \ref{fig_ionization} (b,c).

\begin{figure}[!ht]
  \includegraphics[width=\linewidth]{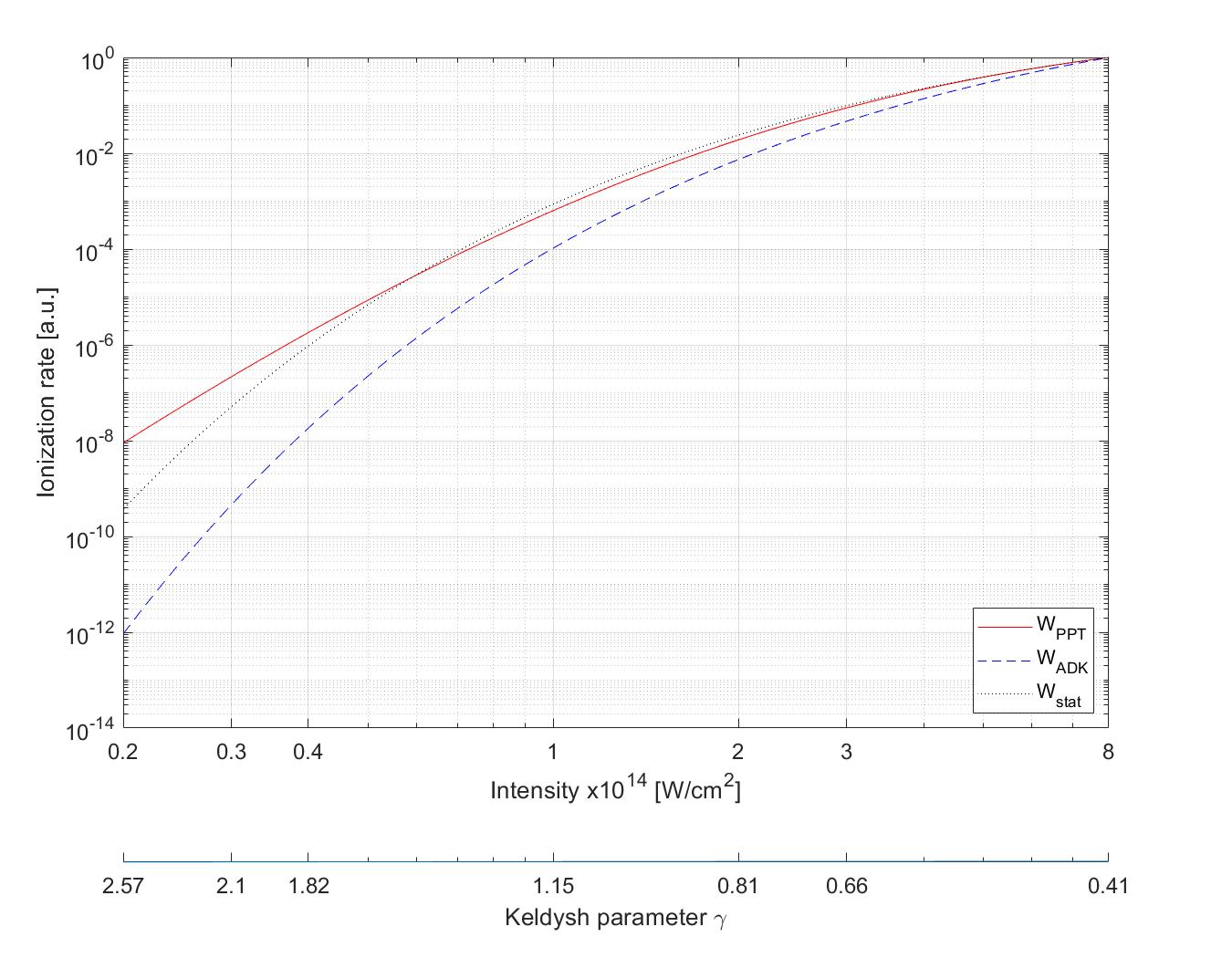}
  \caption{\label{Fig_ion_rate} Ionization rate for Ar ($I_p=15.76eV$) in 800nm linearly polarised laser field, calculated using the ADK, shown as a dashed blue line, and PPT, shown as a solid red line, strong field ionization models. Ionization rate of the Hydrogen atom ($I_p=13.6 eV$) in the static electric field, $E=\sqrt{2I/{c \varepsilon_0}}$, is shown as a dotted black line, for reference.}
  %Calculated using :G:\BGU_since_2013\papers\Gas_Jet_ionization_Tomography\Numerical_evaluations_Gas_Jet_ionization\main_ionization_run_v1.m

\end{figure}

Keldysh's expression for the ionization rate successfully captured a key aspect of the phenomenon—the exponential dependence of the ionization probability on the driving field amplitude. However, the exponential pre-factor was incorrect, as its asymptotic behavior ($\omega \rightarrow 0$) did not align with the well-known equation for the tunnel ionization of a hydrogen atom in a constant electric field \cite{Landau_QM_V3} (see Eq. (\ref{Landau_ionization})). This discrepancy arose because Keldysh did not account for the Coulomb interaction between the photoelectron and the parent ion.

Building on Keldysh's groundbreaking work, Perelomov, Popov, and Terentev investigated the impact of Coulomb interaction on ionization rates in their series of papers \cite{Perelomov_PPT_JETP1966, Perelomov_PPT2_JETP1967, Perelomov_PPT3_JETP1967}. Assuming a short-range potential, they derived Eq. (\ref{PPT_ionization}) (see Appendix) for the optical cycle-averaged ionization rate $W_{PPT}(E_L, \omega_L)$ under a linearly polarized driving laser field.
Throughout the rest of this paper, unless explicitly stated otherwise, we will use atomic units ($\hbar = e = m_{\textrm{e}} = 1$).

%[?? look also into the \cite{Lai_SFI_experimentalPRA_2017} Appendix]:
%The formula was taken in the form as appears in the Lai_Experimental_investigation_strong_field_ionization_theories_laser_ fields_from_visible_to_midinfrared_frequencies_PRA_2017:
%Later consider For the circular polarized field and wavelength depencdence on the field {?Make calculations for both linear and circular polarisation cases??)

 %across the beam we go from tunnelling ($\gamma = ? typical HHG gamma?$ via intermediate to multiphoton ionization ($\gamma=?typical HHG gamma?$) regimes as the Gaussian beam strength vanishes rapidly across the beam as shown in Fig \ref{fig:ionization}. (See Fig. ?? show cross section of the Gaussian beam?? Note also that in the HHG most electrons do not return to the parent ion and do not participate in the HHG. Those free ions are detected by the ion detector as depicted in the figure ???.

 %[?? Mention the problems in Keldysh theory and justify usage of PPT and ADK ?? Note that Keldysh's theory neglects the Coulomb interaction of the freed electron and couples the bound states of the parent atom/ molecule the free space Volkov states??]
%[?? Mention what Keldysh parameter in HHG ??]

Yet another often-used ionization model is the Ammosov, Delone, Krainov (ADK) model Eq. (\ref{ADK_ionization_rate}), derived from the PPT model in the limit as $\gamma \rightarrow 0$.  While the ADK model offers simplicity, its applicability is confined to the tunneling regime with $\gamma \leq 0.5$, as confirmed experimentally  \cite{Ilkov_Ionization_atoms_tunnel_J_Phys_B_1992}. 
Figure \ref{Fig_ion_rate} illustrates the cycle-averaged ionization rate, as predicted by both the PPT and ADK models, plotted against driving laser intensity and the corresponding Keldysh parameter $\gamma$.

Continued improvement of strong-field ionization models, coupled with rigorous experimental validation, remains a dynamic area of research  \cite{Reiss_effect_PRA_1980, Ammosov_ADK_JETP_1986, Yudin_Ivanov_ion_PRA_2001, Becker_ionization_rates_PRA_2001, Tong_LinCD_Molecular_ADK_PRA2002, Popruzhenko_ionization_arbitrary_laser_PRL_2008, Zhao_intensity_ion_calib_PRA_2016, Larochelle1_Coulomb_ionization_J_Phys_B_1998, Cornaggia_ionization_molecules_PRA_2000, Wu_Dorner_Multiorbital_Tunneling_Ionization_of_the_CO_Molecule_PRL_2012, Lai_SFI_experimentalPRA_2017, Lloyd_Comparison_strong_field_ionization_models_OpExp_2019}. Experimental evidence shows that the ionization rate predicted by the PPT model is highly accurate for atoms up to a Keldysh parameter of $\gamma \sim 3-4$ \cite{Larochelle1_Coulomb_ionization_J_Phys_B_1998, Lai_SFI_experimentalPRA_2017} and remains valid even for small molecules \cite{Cornaggia_ionization_molecules_PRA_2000}.
In the following analysis, we will use the PPT model for the ionization rate Eq. (\ref{PPT_ionization}). 

While different models will naturally introduce quantitative differences, the key findings presented in this work remain unaffected. In this study, we focus on laser-induced single ionization. It is important to note, however, that under certain conditions, double ionization may play a role \cite{Shlomo_In_situ_arXiv_2024}, which will be a subject for future studies.
For the sake of brevity, we will employ the term ``atom'' throughout the remainder of this paper, although our findings are equally pertinent to molecules.

%For future work:
%??It will be interesting to compare the
%For example for the laser intensities typical for the HHG  of???, the condition of $\gamma = 4$ encircles ??? of laser power and the beam intensity for the $\gamma=3$ is only ?? percent when compared to the maximum value
%[??Fig. \ref{fig_ionization} b.??].
%We will develop our further treatment using the PPT model for the ionization rate [?? justify why??] as appears in Eq. ??.
%[??However, we can easily incorporate these new model in our treatment??]
%[??However, while different model will obviously have quantitative influence, the key findings will outlined in this work will remain unchanged??]

\subsection{Density of the laser induced strong-field Ionized atoms within the medium in space and time}
\label{Section_Density}

In the previous section, we discussed the ionization rate in the context of relevant strong field ionization models. Now, we turn our attention to the total probability of ionization, $P_i(\overrightarrow{\textbf{r}})$, (and hence the ion's density) by the laser pulse at any specific location, $\overrightarrow{\textbf{r}}=(x,y,z)$, within the measured medium.

The total probability of ionization at a point $\overrightarrow{\textbf{r}}$ and time $t$ is a functional of the electric field's time evolution, $E_L(\overrightarrow{\textbf{r}},t)$, within a laser pulse during the time period $(-\infty,t]$, i.e., $P_i(\overrightarrow{\textbf{r}},t)=P_i[E_L(\overrightarrow{\textbf{r}},t)]$. In principle, the ionization rate, $W_i(\overrightarrow{\textbf{r}},t)$, depends on space and time via the space-time dependence of the electric field $E_L(\overrightarrow{\textbf{r}},t)$. For the sake of simplicity, we will often drop the explicit symbol of location, $\overrightarrow{\textbf{r}}$, from the total ionization probability, $P_i$, and the ionization rate, $W_i$, expressions, tacitly assuming their coordinate dependence. The derivation of the expression for total ionization probability, Eq. (\ref{prob_ion5}), as a function of ionization rate appears in Appendix \ref{Appendix_probability}.

%$P_i(x,y,z)=P_i[E_L(x,y,z,t)]$ - probability of ionization after the pulse was gone. The $P_i(x,y,z)$ is the functional of the time evolution of the laser field.
%
%$E_L(x,y,z,t)$ at any given point of space $(x,y,z)$. (?? Make a notation similar to the notation of action as the functional of Lagrangian??)

 Note that most of the theoretical ionization models discussed in the previous sections, such as ADK or PPT, provide ionization rates, $W_i$, averaged over a half-optical cycle. Since we are interested in the total probability of ionization by the time the ionizing laser pulse has passed, i.e., as $t \rightarrow +\infty$, the cycle-averaged ionization rate is suitable for our calculations. Using Eq. (\ref{prob_ion5}), we can calculate the total ionization probability within the medium. If the measured density distribution does not change significantly during the passage of the laser pulse, we can write:

\begin{equation}
\label{prob_ion_after_pulse}
P_i(\overrightarrow{\textbf{r}}) = 1 - \exp \left[ {- \int_{-\infty}^{+\infty} W(\overrightarrow{\textbf{r}},t') dt'} \right]
\end{equation}

\begin{figure}[!ht]
  \includegraphics[width=0.7 \linewidth]{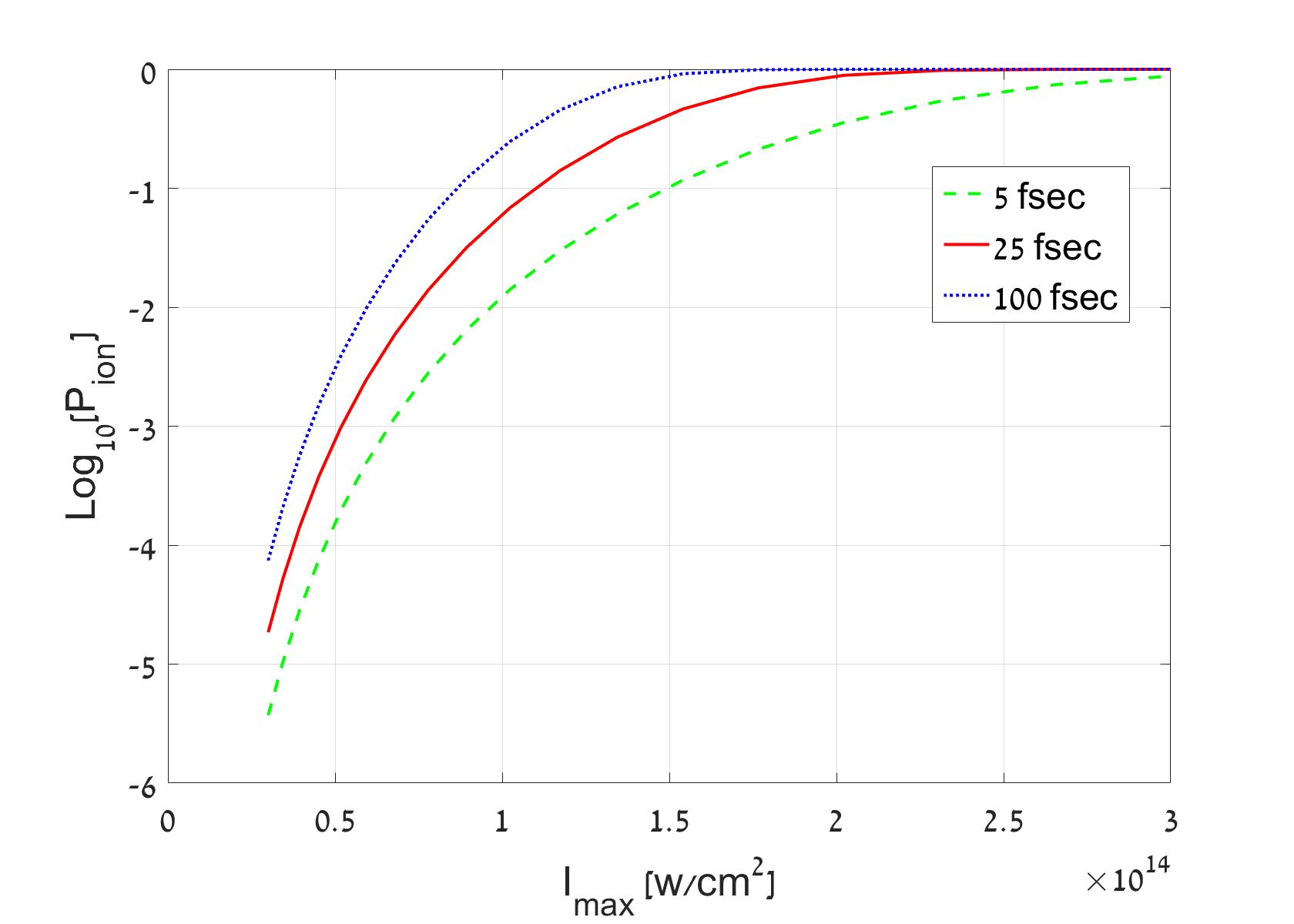}
  \caption{ \label{Fig_ion_prob} Ionization probability for Ar ($I_p=15.76eV$) in 800nm linearly polarised laser field for Gaussian temporal profile pulse, $I(t)=I_{max} \exp\{-2(t/\tau_G)^2\}$.  Full width half maximum pulse duration of the intensity profile ($\tau_{FWHM}=\sqrt{2 \ln{2}}\tau_G$) is shown in the figure's legend.}
  %Calculated using :G:\BGU_since_2013\papers\Gas_Jet_ionization_Tomography\Numerical_evaluations_Gas_Jet_ionization\main_ionization_vs_wavelength_and_intensity_v1.m

\end{figure}

In many cases of interest, the time-scale of the dynamics of the phenomena of interest is much longer than the duration of the laser pulse. For example, the typical gas pulse duration in the medium used for high harmonic generation (HHG) is a few tens of microseconds ($10^{-6}$ $\textrm{sec}$) \cite{Even_Even_Lavie_valve_EPJ2015even} or even longer \cite{Frumker_orientation_HHG_PRL2012}, while the ultrafast laser pulse duration is on the order of a few tens of femtoseconds ($10^{-15}$ $\textrm{sec}$). This means the laser pulse duration is 9 orders of magnitude shorter than the time-scale of the relevant target dynamics involved.

In such a case, the measured medium appears effectively "frozen" during the ionizing pulse. Then we can write for the density of ionized atoms $n_i(\overrightarrow{\textbf{r}},t)$, at the position $\overrightarrow{\textbf{r}}$, and at the time $t$, within the medium with neutrals density $n_0(\overrightarrow{\textbf{r}},t)$:

\begin{equation}
\label{density_ionization_t}
 n_i(\overrightarrow{\textbf{r}},t)=P_i(\overrightarrow{\textbf{r}}) \times n_0(\overrightarrow{\textbf{r}}, t)
 \end{equation}

Considering dynamic tomography \cite{Shlomo_4D_tomography_arXiv_2024}, we have to take into account both the time it takes for the laser pulse to cross the measured region of interest, as well as the pulse duration (whichever is longer). For example, in the discussed case of the gas pulsed jet, the extent of the typical region of interest, $L_{jet}$, is about a millimeter or less ($L_{jet}=1 \textrm{mm}$). In such a case, a relevant time-scale to compare with would be $L_{jet}/c\simeq 3$ picoseconds, which is still much faster (~6 orders of magnitude) than the measured physical phenomena, and the approximation (\ref{density_ionization_t}) works very well.
We can follow the time evolution dynamics of the spatial density distribution $n_0(\textbf{r},t)$ by scanning the time delay between the laser pulse hitting the target and the event that initiates the dynamics.

While the femtosecond pulse duration sets the fundamental limit on the temporal resolution in studying the dynamics, in reality, we are usually limited by the electronics jitter of the time delay generator, which is on the order of tens of picoseconds for state-of-the-art systems \cite{Shlomo_4D_tomography_arXiv_2024}.

\section{Laser Induced Strong-Field Ionization Tomographic imaging}
\label{Section_Tomography}

\subsection{Measurement and Recostruction Principles}
\label{Section_TPrinciples}

\begin{figure}
%Tentative plots:
%"G:\NRC_work_since_December2007\PostDoc_Papers\wavefront_hhg\Set_up_solidworks_figure\Set_up_assembly.SLDASM"
%"G:\BGU_since_2013\papers\Gas_Jet_ionization_Tomography\filesOfArticle\Fig_Setup_v1.jpg"
%"G:\BGU_since_2013\papers\Gas_Jet_ionization_Tomography\Numerical_evaluations_Gas_Jet_ionization\Figures_making4Theory_Gas_Jet_ionization_paper.pptx"
    \begin{center}
\includegraphics[width=0.7\linewidth]{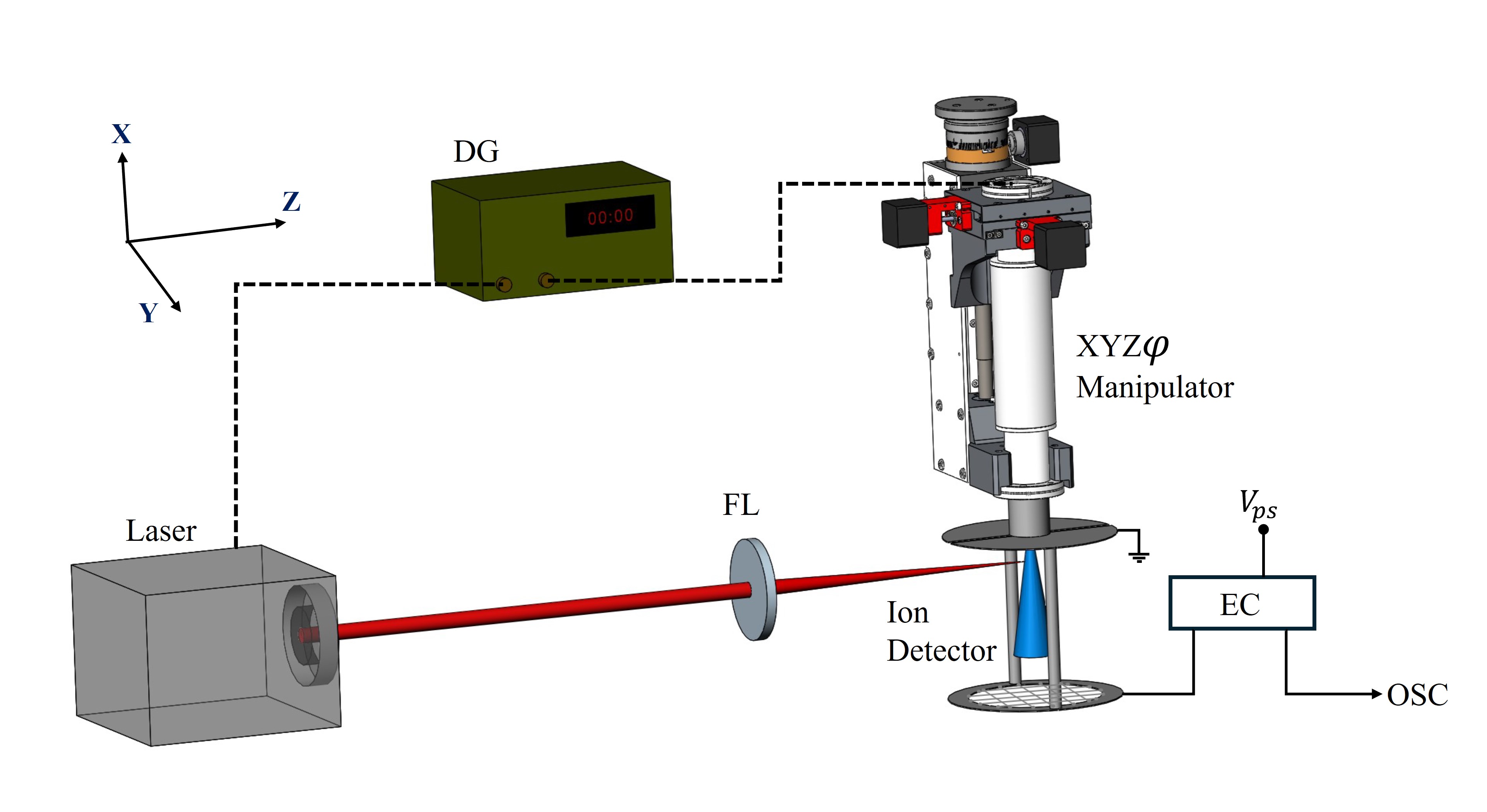}

    \caption{\label{fig_system} A schematic representation of the experimental setup for the LISFIT.}

    \end{center}

\end{figure}	

With the necessary tools to calculate strong-field ionization in the medium, we now turn our attention to laser-induced strong-field ionization tomographic imaging. Consider a laser beam focused by a lens (FL) into the measured medium, as illustrated in Fig. \ref{fig_system}. For clarity, we will examine the density distribution within a pulsed gas jet \cite{Tchulov_LaserStrongFieldTom_ScRep_2017, Shlomo_4D_tomography_arXiv_2024}, although our analysis applies to any form of LISFIT and is relevant for understanding key aspects of strong-field metrology in general.

The measured target is positioned on an $XYZ \varphi$ manipulator, enabling 3D scanning and rotation around a fixed axis. First, the Z and Y positions are adjusted so that the laser beam is focused close to the rotational axis. For each height (X), we scan along the Y direction across the laser beam and, in cases without rotational symmetry, allow rotational scanning as well. For each position, the strong-field ionization signal is measured using an ion detector (ID) and accompanying electronics. Without delving into the experimental implementation details, which are available elsewhere \cite{Tchulov_LaserStrongFieldTom_ScRep_2017, Shlomo_4D_tomography_arXiv_2024}, it is worth noting that the entire apparatus is compact and lightweight, making it suitable for in situ measurements.

 For instance, in high-harmonic generation (HHG), most laser-ionized electrons do not return to recombine with the parent ion and, therefore, do not contribute to the HHG process. These free electrons and ions are detected by the ion detector (ID) in situ \cite{Tchulov_LaserStrongFieldTom_ScRep_2017, Shlomo_4D_tomography_arXiv_2024}.

Recently, we extended LISFIT into the fourth, temporal dimension by controlling the arrival of the ultrafast ionizing laser pulse relative to the triggering timing (in this case, the gas injection timing \cite{Shlomo_4D_tomography_arXiv_2024}). This was achieved by using a digital delay/pulse generator (DG) shown in Fig. \ref{fig_system}. Scanning both the trigger timing relative to the ionizing laser pulse across the time window of interest, as well as the laser beam position relative to the measured target, allows us to capture a complete picture of the process in both space and time.

The number of atoms $N_i(x_0,y_0)$, ionized by the laser beam crossing the medium at the point $\overrightarrow{\textbf{r}_{0 \perp}} = (x_0,y_0)$ (see axis orientation defined in Fig. \ref{fig_system}) along the Z axis is given by

\begin{multline}
\label{N_i}
N_i(\overrightarrow{\textbf{r}_{0 \perp}})=N_i(x_0,y_0)=\iiint_V {P_i(x-x_0,y-y_0,z)n_0(x,y,z)} \,dx\,dy\,dz = \\
= \int_{- \infty}^{+ \infty} \,dz \iint_{S_{\perp}(z)} {P_i(x-x_0,y-y_0,z)n_0(x,y,z)} \,dx\,dy
\end{multline}

where the double integration is over entire $S_{\perp}(z)$ plane, perpendicular to the laser beam propagating along Z-axis at some distance \textit{z} along the beam.
In this expression the integration is done over the entire space, but effectively it is limited  to the focal volume - the region in space where the $P_i(x-x_0,y-y_0,z)$, given by Eq. (\ref{prob_ion_after_pulse}) doesn't vanish.

The double integral, $\iint_{S_{\perp}(z)} {P_i(x-x_0,y-y_0,z)n_0(x,y,z)} \,dx\,dy$, constitutes a correlation of the spatial neutrals density distribution, $n_0(x,y,z)$, and the laser-induced ionization probability distribution, $P_i(x,y,z)$, in the plane $S_{\perp}(z)$. Thus, the spatial density distribution is averaged around the point $(\overrightarrow{\textbf{r}_{0 \perp}},z) = (x_0,y_0,z)$ in the $S_{\perp}(z)$ plane with the weight factor given by $P_i(x,y,z)$, leading to the loss of spatial resolution and apodization.

Then we can write:

\begin{multline}
\label{lamda_Ni}
\overline{\lambda}_{N_i}(\overrightarrow{\textbf{r}_{0 \perp}},z) =\left( P_i \star n_0 \right) (x_0,y_0) = \iint_{S_{\perp}(z)} {P_i(x-x_0,y-y_0,z)n_0(x,y,z)} \,dx\,dy=\\
= \overline{n}_0(\overrightarrow{\textbf{r}_{0 \perp}},z) \times \iint_{S_{\perp}(z)} {P_i(x-x_0,y-y_0,z)} \,dx\,dy = \overline{n}_0(\overrightarrow{\textbf{r}_{0 \perp}},z) \times \sigma_i(z)
\end{multline}

where the $\overline{\lambda}_{N_i}(\overrightarrow{\textbf{r}_{0 \perp}},z)$ is the number of ionized atoms per unit length along the laser beam   propagation (or linear ions density), and $\overline{n}_0(x_0,y_0,z)$ is the neutrals density averaged across the $S_{\perp}(z)$ plane  within the laser beam, crossing the medium at the point $\overrightarrow{\textbf{r}_{0 \perp}} = (x_0,y_0)$, and ionization cross-section $\sigma_i(z)$. 

For the sake of clarity and simplicity, we will assume an undepleted pump and neglect beam distortions within the measured medium in our analysis. Note that the laser-induced ionization profile across the $S_{\perp}(z)$ plane  is independent of  $\overrightarrow{\textbf{r}_{0 \perp}}$, therefore the ionization cross-section $\sigma_i(z)$ in Eq. (\ref{lamda_Ni}) can be defined as

\begin{equation}
\label{sigma_ion}
\sigma_i(z)  \triangleq \iint_{S_{\perp}(z)} {P_i(x,y,z)} \,dx\,dy
\end{equation}

%See the page 6 of the Siegman paper

Next, let's define the width of the laser-induced ionization probability distribution $P_i(x,y,z)$ across the $S_{\perp}(z)$ plane. The spatial distribution of ionization probability depends on the spatial distribution of the driving laser beam's intensity. Hence, similar to the ambiguity in defining the laser beam width, there is no single universally meaningful definition of the ionization profile width.

Fundamentally, this uncertainty stems from the fact that there is no unique rule for estimating the product of the near-field and far-field widths for an arbitrary laser beam. However, it is well known that "naive" definitions of width, such as full width at half maximum (FWHM), $1/e^2$ width, etc., have notable shortcomings in the context of laser beam characterization \cite{siegman1998maybe}.

We will adopt a definition analogous to the commonly accepted laser beam width definition used widely for laser beam quality characterization, in the so-called "M-squared" method, which is the variance or second moment width \cite{siegman1998maybe}. Using the second moment method, we define the radius of the ionization, $R_{SM}(z)\triangleq \sqrt{<r^2>}$. For simplicity, we will assume a round beam profile (meaning $<x^2>=<y^2> \Rightarrow R_{SM}(z) =\sqrt{<x^2>+<y^2>} =\sqrt{2<x^2> }$  that is independent of the specific crossing point, $\overrightarrow{\textbf{r}_{0 \perp}} = (x_0,y_0)$, so that

\begin{equation}
\label{r_sigma_2moment}
R_{SM}(z)=\sqrt{\frac{2\iint_{S_{\perp}(z)} {x^2}{P_i(x,y,z)} \,dx\,dy}{\iint_{S_{\perp}(z)} {P_i(x',y',z)} \,dx'\,dy'}}
\end{equation}

and the corresponding ionization cross-section area $\sigma_{SM}(z)$:

\begin{equation}
\label{sigma_via_second_moment}
\sigma_{SM}(z)=\pi R_{SM}(z)^2
\end{equation}

%[? Discuss that $\sigma_{SM}(z)\neq \sigma_{i}(z)$, because $P_i(x,y,z)$ can be less than 1. Provide example that the same $\sigma_{i}(z)$ can be obtained with different ionization width $R_{SM}(z)$. May be show plot. Discuss meaning of $\sigma_{SM}(z)$ and $ \sigma_{i}(z)$ in the context of tomography. While $\sigma_{SM}(z)$ influences the spatial, resolution of the measurement; the $ \sigma_{i}(z)$ impact the overall signal (apodization, localization). Try to make analogy with tomographic reconstruction from partially available info (only some angles provide not apodalzed measurement)?]

Substituting  expression (\ref{lamda_Ni}) into Eq. (\ref{N_i}), we obtain:

\begin{equation}
\label{N_i_line_integral}
N_i(\overrightarrow{\textbf{r}_{0 \perp}})= \int_{- \infty}^{+ \infty} \overline{\lambda}_{N_i}(\overrightarrow{\textbf{r}_{0 \perp}},z) \,dz
\end{equation}

\begin{figure}
%"G:\BGU_since_2013\papers\Gas_Jet_ionization_Tomography\Numerical_evaluations_Gas_Jet_ionization\Figures_making4Theory_Gas_Jet_ionization_paper.pptx"
%"G:\BGU_since_2013\papers\Gas_Jet_ionization_Tomography\Numerical_evaluations_Gas_Jet_ionization\main_tomography_simulation_v2.m“
    %\begin{center}
	%\includegraphics[width=0.7\linewidth]{Figure_Tomography_General_v2.jpg}
\includegraphics[width=0.7\linewidth]{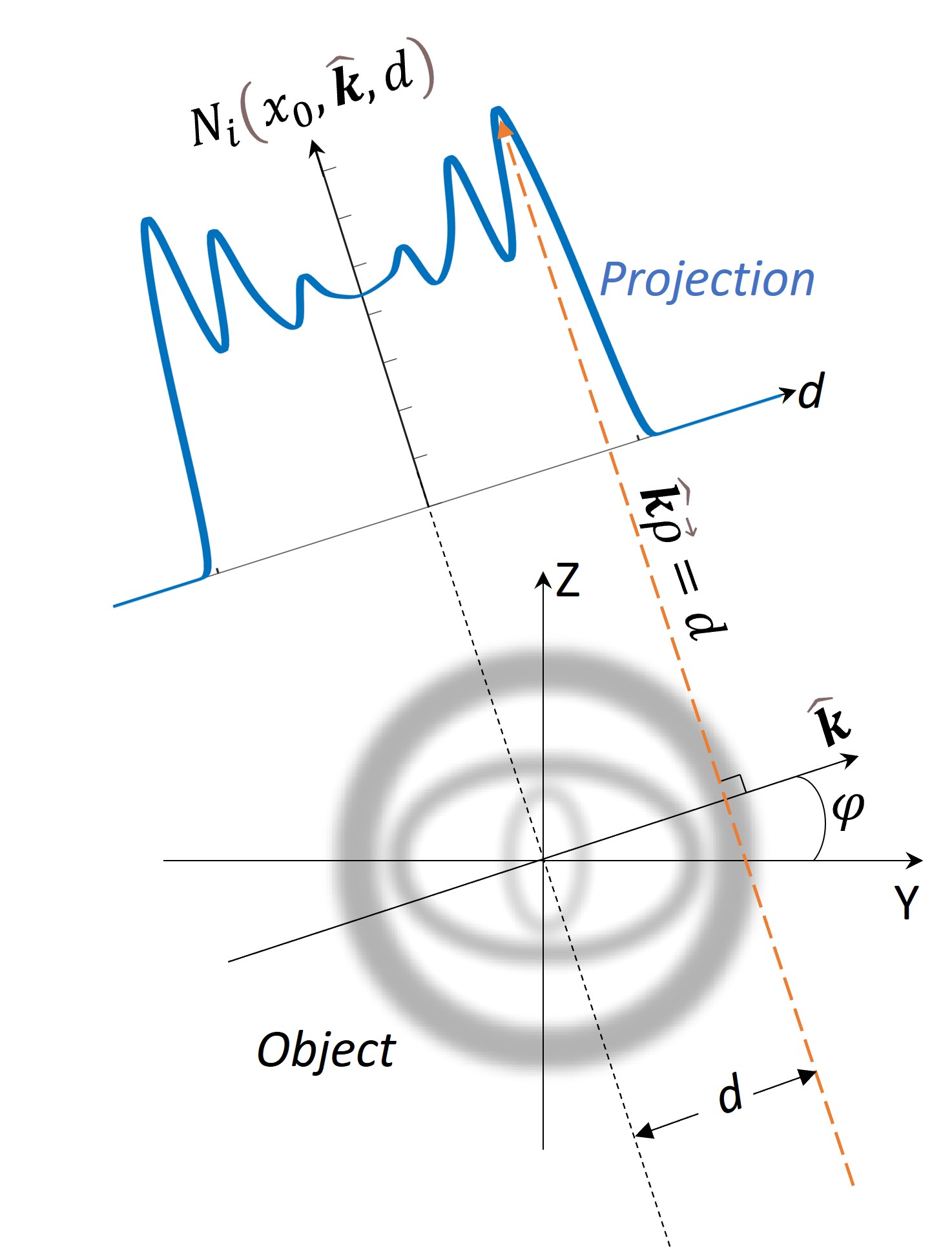}

    \caption{Laser indused strong-field interaction parallel projections tomography. Scanning the distance ($d$) of the laser beam (represented by the dashed red ray) and measuring the resulting ion signal ($N_i$) for a given perpendicular direction vector ($\mathbf{\hat{k}}$) constitute a parallel projection.}

    \label{Fig_Tomography_General}
    %\end{center}
\end{figure}

The line integral in Eq. (\ref{N_i_line_integral}), often referred to as a \textit{ray integral}, constitutes a basic building block of computerized tomography \cite{Born_wolf_principles_optics_1999}. In the general case, we can measure this line integral for arbitrary orientation of the driving laser beam relative to the medium, as shown in Fig. \ref{Fig_Tomography_General}. This laser beam, which corresponds to the ray in its mathematical interpretation in the context of tomography, is defined by vector $\mathbf{\hat{k}}$ perpendicular to this beam and its distance $d$ from the origin, as illustrated in Fig. \ref{Fig_Tomography_General}.

Let's first consider the case where the extent of the measured medium (e.g., gas jet) in the (YZ) plane is much smaller than the Rayleigh range of the driving laser beam. For example, these conditions were very well satisfied in our first proof-of-principle experiments for static \cite{Tchulov_LaserStrongFieldTom_ScRep_2017} and 4D dynamic \cite{Shlomo_4D_tomography_arXiv_2024} strong field tomography. In such a case, both $\sigma_{i}$ and $\sigma_{SM}$ remain constant in the region of interest. The projection corresponding to the generated total ionization signal is then equal to the line integral along this laser beam:

\begin{equation}
\label{Ni_general}
N_i (x_0,\mathbf{\hat{k}},d) = \iint_{S_{yz}} {\overline{\lambda}_{N_i}(x_0,y,z)} \delta (d-\mathbf{\hat{k}} \vec{\rho}) \,dy\,dz
\end{equation}

where $\vec{\rho}= z \mathbf{\hat{e}_z} + y \mathbf{\hat{e}_y}$
is a point in the object at the height $x_0$, and this double integral with the delta function, $\delta (d-\mathbf{\hat{k}} \vec{\rho}$), defines the line integral, $\int_{\mathbf{\hat{k}} \vec{\rho}=d}{{\overline{\lambda}_{N_i}(x_0,y,z)}} d l=\iint_{S_{yz}} {\overline{\lambda}_{N_i}(x_0,y,z)} \delta (d-\mathbf{\hat{k}} \vec{\rho}) \,dy\,dz$  along the laser beam $\mathbf{\hat{k}} \vec{\rho}=d$. In this general case, the double integral in the Eq. (\ref{lamda_Ni}) should be taken over the  plane ${S_{\perp}(l)}$  perpendicular to the direction defined by the ray $\mathbf{\hat{k}} \vec{\rho}=d$.

Scanning the laser beam by varying  the distance $d$  and measuring the ion signal $N_i (x_0,\mathbf{\hat{k}},d)$ for a specific $\mathbf{\hat{k}}= \cos \varphi  \mathbf{\hat{e}_z} + \sin \varphi \mathbf{\hat{e}_y} $ , as schematically shown in Fig. \ref{Fig_Tomography_General}, forms a \textit{ parallel projection}.
The set of parallel projections for all possible direction of  $\mathbf{\hat{k}}$ is called  a \textit{ Radon transform}.

%% Complete this derivation which is central to this paper.!!!

The central theorem of the computed tomography \cite{Born_wolf_principles_optics_1999,Kak_book_tomography_2002}, called  \textit{ the projection-slice theorem or the Fourier slice theorem}, applied to our problem states \footnote{For convenience in computer implementation, here we adopt a unitary, ordinary (not angular) frequency convention for the Fourier Transform definition.}:

\begin{equation}
\label{Fourier_lamda_Ni}
\tilde{\overline{\lambda}}_{N_i}(x_0,kk_y,kk_z)= \int^{+\infty}_{-\infty} {N_i (x_0,\mathbf{\hat{k}},d) \exp{\left[-i 2 \pi k d \right]}} \,\textrm{d}d \triangleq \tilde{N}_i (x_0,\mathbf{\hat{k}},k)
\end{equation}

 It shows that the two-dimensional Fourier transform $\tilde{\overline{\lambda}}_{N_i}(x_0,kk_y,kk_z)$  of the linear ion density  ${\overline{\lambda}_{N_i}(x_0,y,z)}$ with respect to $\vec{\rho}$ can be found by taking a one-dimensional Fourier transform  $\tilde{N}_i (x_0,\mathbf{\hat{k}},u)$ with respect to the scalar parameter $d$ of the total number of atoms, $N_i (x_0,\mathbf{\hat{k}},d)$, ionized by the laser beam propagating along the ray $\mathbf{\hat{k}} \vec{\rho}=d$.
Taking the inverse Fourier transform of the $\tilde{\overline{\lambda}}_{N_i}(x_0,f_y,f_z)$ , where $f_y=kk_y$ and $f_z=kk_z$, facilitates the reconstruction of the two-dimensional distribution of the ${\overline{\lambda}_{N_i}(x_0,y,z)}$:

\begin{equation}
\label{Inverse_Fourier_lamda_Ni}
\overline{\lambda}_{N_i}(x_0,y,z)= \iint{\tilde{\overline{\lambda}}_{N_i}(x_0,f_y,f_z) \exp{\left[i 2 \pi (f_y y +f_z z)\right]}} \,d f_y\,d f_z
\end{equation}

Repeating this measurement and reconstruction procedure for different heights $x_0$, as shown in Fig. \ref{fig_system}, would enable the three-dimensional reconstruction of ${\overline{\lambda}_{N_i}(x,y,z)}$ in the region of interest.

%[??See the derivation in the Notability file "Theory jet tomography 2021"??]

The averaged density of ionized atoms, $\overline{n}_i(\overrightarrow{\textbf{r}})$ averaged within an ionization focal spot, can be expressed as:

\begin{equation}
\label{n_i_recostructed}
\overline{n}_i(\overrightarrow{\textbf{r}})=\frac{\overline{\lambda}_{N_i}(\overrightarrow{\textbf{r}})}{\sigma_i}
\end{equation}

Given the spatial distribution of the ionized atoms, the averaged neutrals density distribution $\overline{n}_0(\overrightarrow{\textbf{r}})$ within the sample can be derived using Eq. (\ref{density_ionization_t}):  

\begin{equation}
\label{n_0_recostructed}
\overline{n}_0(\overrightarrow{\textbf{r}})=\overline{n}_i(\overrightarrow{\textbf{r}})/P_i(\overrightarrow{\textbf{r}})
\end{equation}

Note that in practice, the projections $N_i (x_0,\mathbf{\hat{k}},d)$ are measured not across a continuous, but a discrete set of angles, always accompanied by some level of noise. It turns out that in this case, direct application of Eq. (\ref{Inverse_Fourier_lamda_Ni})  produces noisy results with artifacts \cite{Kak_book_tomography_2002, Deans_book_Radon_2007}.
Equation (\ref{Inverse_Fourier_lamda_Ni}) in polar coordinates can be written as:

%See the derivation in my Ipad notes fro this derivations "Theory jet tomography".
\begin{equation}
\label{Polar_Inverse_Fourier_lamda_Ni}
\overline{\lambda}_{N_i}(x_0,y,z) = \int_{0}^{\pi} \,d \varphi \int_{-\infty}^{+\infty} |k| \tilde{\overline{\lambda}}_{N_i}(x_0,k \sin{\varphi},k \cos{\varphi}) \exp{\left[i 2 \pi k (y\sin{\varphi}+ z \cos{\varphi}) \right]} \,d k
\end{equation}

Application of this formula to the tomographic reconstruction with a discrete set of sampled projections in the  real-world scenario produces much better results than  Eq. (\ref{Inverse_Fourier_lamda_Ni}). This is by far the most often used reconstruction approach, dubbed a \textit{filtered backprojection algorithm}, for the practical applications \cite{Kak_book_tomography_2002}.

From an experimental standpoint, we have recently demonstrated \cite{Shlomo_In_situ_arXiv_2024, Shlomo_4D_tomography_arXiv_2024} that by measuring the laser beam profile, total charge generated, and applying the presented theoretical approach, we obtain the laser-generated ion density and absolute gas density distribution in the interaction region.

\subsection {Axial symmetry, Abel Transform}
\label{Section_Axial_Symmetry}

An important special case is a tomographic measurement of a medium possessing axial symmetry (AS).
A pulsed gas jet, often used in HHG  and attosecond science experiments \cite{Frumker_orientation_HHG_PRL2012, Itatani_tomog_Nature2004},
is an example of such a medium.
In this case, the ion density depends only on $\rho=\sqrt{y^2+z^2}$, so that $\overline{\lambda}_{N_i} (x_0,y,z)=\overline{\lambda}_{N_i}^{AS} (x_0,\rho=\sqrt{y^2+z^2})$. Hence, the projections in Eq. (\ref{Ni_general}) are independent of the direction $\mathbf{\hat{k}}$, and therefore it is enough to measure only one set of projections, $N_i^{AS} (x_0,d)=N_i (x_0,\mathbf{\hat{k}},d)\: \forall \mathbf{\hat{k}}$, i.e., for arbitrary angle $\varphi$ in Fig. \ref{Fig_Tomography_General}.
%See derivation in my Notability notes: "Theory jet tomography 2021".
%The derivation also appears in G:\evgeny\books_in_files\books_tomography\Stanley_Deans_Radon_and_Abel_Transform.pdf
%page 52 of 96, Section 8.11.1
For simplicity and without loss of generality, let's assume $\mathbf{\hat{k}}=\mathbf{\hat{y}}$. Then, using Eq. (\ref{Ni_general}), we have:

\begin{multline}
\label{Ni_r_rot_symmetry}
N_i^{AS} (x_0,d) = \iint_{S_{yz}} {\overline{\lambda}_{N_i}^{AS} (x_0,\sqrt{y^2+z^2})} \delta (d-y) \,dy\,dz= \int^{+\infty}_{-\infty} {\overline{\lambda}_{N_i}^{AS} (x_0,\sqrt{d^2+z^2})} d z \\
= 2 \int^{+\infty}_{0} {\overline{\lambda}_{N_i}^{AS} (x_0,\sqrt{d^2+z^2})} d z
\end{multline}

Substituting $r = \sqrt{d^2+z^2}$, and noting that $z = \sqrt{r^2-d^2} \Rightarrow dz=rdr/\sqrt{r^2-d^2}$, we have:

\begin{equation}
\label{Ni_r_Abel}
N_i^{AS} (x_0,d) =  2 \int^{+\infty}_{|d|} \frac{\overline{\lambda}_{N_i}^{AS} (x_0,r) r}{\sqrt{r^2-d^2}} d r
\end{equation}

This is called an {\textit{Abel transform}. In its analytical form, the inverse of the Abel transform is given by: $\overline{\lambda}_{N_i}^{AS} (x_0,r) = -\frac{1}{\pi} \int^{+\infty}_{r} {N_i^{AS}} ' (x_0,\xi) (\xi^2-r^2)^{-1/2} d \xi$ \cite{Bracewell_Fourier_book_1986} %\cite{Bracewell_Fourier_book_1986}.

%???
%[?This paragraph should be improved?] Another approach to address the axially symmetric medium is to use the Radon transform. For each $\mathbf{\hat{k}}$,  one can either use the same angle-independent parallel projection measurement or improve the signal-to-noise ratio (SNR) by using an averaged  parallel projection from a sample set of parallel projection  measurements. Alternatively, for each $\mathbf{\hat{k}}$, projections from a set of repetitively measured parallel projections  can be utilized \cite{Shlomo_4D_tomography_arXiv_2024}.
%???

 Another approach to address the axially symmetric medium is to use the inverse Radon transform Eq. (\ref{Inverse_Fourier_lamda_Ni}) in  a scenario where the projections are identical for different directions $\mathbf{\hat{k}}$ due to the axial symmetry, except for measurement noise.  Hence, for each $\mathbf{\hat{k}}$ in the inverse Radon reconstruction,  one can either use the same angle-independent parallel projection measurement or, to improve the signal-to-noise ratio (SNR), use a set of independantly measured  parallel projections for the same $\mathbf{\hat{k}}$. This set of redundant projections can be either averaged prior to applying the inverse Radon transform  or, for each $\mathbf{\hat{k}}$ in the inverse Radon transform, various projections from this measured set can be utilized  \cite{Shlomo_4D_tomography_arXiv_2024}.

%\textbf{\textit{[?? End of the alternative derivation. Probably can be dropped. Check that every aspect is covered??}}

%??Discuss the issue of undepleted pump. Is it valid for typical gas target? If not derive the expression for depleted pump. - analogues to how it is treated for absorbtion tomography (see Born and Wolf)Also discuss the issue preserving the beam shape - no Space-time propagation effects due to free electrons dispersion??
%
%Thus we can apply directly tomographic theory (see for example Born and Wolf chapter??) taking into account that we reconstruct ${\overline{n_0(x_0,y_0,z)} g(z)}$.
%To get the density $\overline{n_0(x_0,y_0,z)}$ one has to divide the reconstructed entity ${\overline{n_0(x_0,y_0,z)} g(z)}$ by the "apodization" factor $g(z)$.

\subsection{Resolution}
\label{Section_Resolution}

Spatial resolution is one of the most important characteristics for any imaging technique, including tomography \cite{Haddad_Ultrahigh_resolution_Science_1994, Fessler_Spatial_Res_tomographs_1996, Tsekenis_Spatially_resolved_RSI_2015}, and it will be discussed here.

Equation (\ref{lamda_Ni}) implies that the width and shape of the total probability of ionization, $P_i(x,y,z)$, across the laser beam propagation direction ultimately define the spatial resolution limit of strong field ionization tomography. This is somewhat analogous to the role of the point spread function (PSF) in an imaging system: The density, $n_0(x,y,z)$, details finer than a scale defined by Eq. (\ref{r_sigma_2moment}), will be smeared by the convolution in Eq. (\ref{lamda_Ni}), similar to the resolution loss in an imaging system due to a finite width of the PSF. While in strong field ionization tomography, the width and shape of $P_i(x,y,z)$ across the laser beam are defined by the shape of the driving laser beam and the physics of the strong field ionization phenomena as described in Sections \ref{Section_Key_milestones} and \ref{Section_TPrinciples}, the PSF of an imaging system is determined by aberrations and/or diffraction.

However, in the case of tomography, the situation is more complex compared to direct imaging. The eventual spatial resolution (resolution "voxel") is defined by a combination of this convolution and the number of angular projection samples, defined by the vector $\mathbf{\hat{k}}$ in Eq. (\ref{Ni_general}). While the number of projections can be increased using denser angular sampling, the resolution loss due to Eq. (\ref{lamda_Ni}) smearing determines the ultimate spatial resolution of the LISFIT.

\begin{figure}
    %\begin{center}
    %"G:\BGU_since_2013\papers\Gas_Jet_ionization_Tomography\Numerical_evaluations_Gas_Jet_ionization\main_ionization_profiles_v1.m"
    %"G:\BGU_since_2013\papers\Gas_Jet_ionization_Tomography\Numerical_evaluations_Gas_Jet_ionization\Figures_making4Theory_Gas_Jet_ionization_paper.pptx"
	%\includegraphics[width=0.7\linewidth]{Fig_ionization_profile_v1.jpg}
%\includegraphics[width=0.7\linewidth]{Fig_ionization_profile_with_3D_v1.jpg}
\includegraphics[width=1\linewidth]{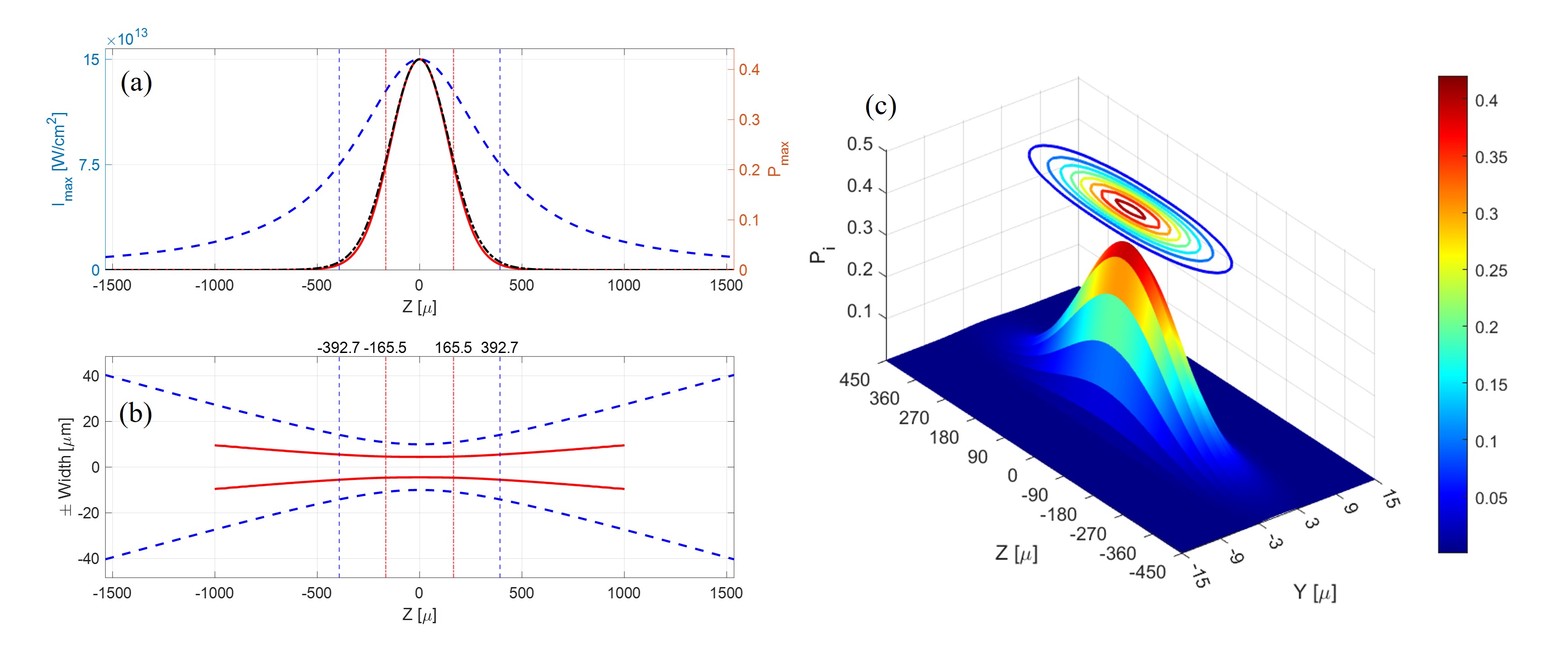}

    \caption{ (a) Laser pulse intensity (dashed blue) and ionization probability (solid red) along the laser propagation axis. The ionization cross section $\sigma_i(z)$, normalized to $P_\text{max}(0)$, is shown by the dash-dotted black curve. (b) Laser beam width, $\pm w(z)$ (dashed blue), and ionization profile width, $\pm W_I(z)$ (solid red), along the beam propagation. (c) Map of ionization probability in argon as a function of position within the laser beam.}

    \label{Fig_ionization_profile_v1}
    %\end{center}
\end{figure}

To realize the full potential of strong-field ionization tomography, it is important to explore the main features of the ionization profile generated by the focused laser beam in the medium. To understand these features, consider a 25 fsec pulse with a Gaussian spatial profile focused to a spot of 10 $\mu m$ that drives the ionization in argon, as shown in Fig. \ref{Fig_ionization_profile_v1}.

The blue dashed curve in Fig. \ref{Fig_ionization_profile_v1} (b) depicts the laser beam width, $w(z)=w_0\times\sqrt{(1+(z/z_R)^2)}$ at the $1/e^{2}$ intensity level, where the Rayleigh range, $z_R=\pi w_0^2/\lambda$, and the central wavelength $\lambda_0=800nm$. The peak intensity of the laser as a function of the distance from the waist, shown as a blue dashed curve in Fig. \ref{Fig_ionization_profile_v1} (a), is proportional to $1/w^2(z)$.

The ionization profile across the laser beam at the waist is shown in Fig. \ref{Fig_ionization_vs_intensity}. Analogous to the definition of laser beam width, we define the ionization width, $w_I$, in terms of the second moment \cite{siegman1998maybe}. Using Eq. (\ref{r_sigma_2moment}), we obtain $w_I(z)=\sqrt{2} R_{SM}(z)$. At the peak laser intensity of $1.5\times10^{14}$ $W/cm^{2}$ at the waist, the ionization width is 4.6 $\mu m$, implying more than a twofold improvement in spatial resolution, facilitated by the nonlinearity of strong-field driven ionization phenomenon.

To characterize the divergence of the ionization profile, it is tempting to define a parameter analogous to the Rayleigh range, $z_R$, where the ionization width, $w_I(z)$, would increase by a factor of $\sqrt{2}$. The solid red curve in Fig. \ref{Fig_ionization_profile_v1} (b) represents $w_I(z)$, and the solid red curve in Fig. \ref{Fig_ionization_profile_v1} (a) depicts the maximum ionization probability, $P_{max}(z)$, as a function of the distance from the waist along the laser beam propagation.

Note that the total ionization probability across the laser beam, shown as a dot-dashed black curve (normalized to the same numerical value as $P_{max}(0)$), has an almost identical profile to $P_{max}(z)$. Also, $P_{max}(z)$ vanishes significantly faster compared to $I_{max}(z)$ due to the strong non-linear nature of the ionization process. The map of ionization probability as a function of position within the laser beam is shown in Fig. \ref{Fig_ionization_profile_v1} (c). The contour lines shown in Fig. \ref{Fig_ionization_profile_v1} (c) above the probability surface plot correspond to the curves of equal probability according to the colorbar scale shown on the right side.

Remarkably, the divergence of the ionization probability is significantly smaller than that of the driving laser beam, as evident from Fig. \ref{Fig_ionization_profile_v1} (b). Notably, at the point where the ionization width increases by a factor of $\sqrt{2}$, the ionization probability drops to a negligible value ($< 10^{-3}$).

Thus, instead of using the conventional "$\sqrt{2} w_0$" criterion to define divergence, we propose characterizing the divergence by measuring the ionization width along the z-axis at the point where $P_{max}(z)$ decreases by a factor of 2. We refer to this position along the z-axis as the \textit{significant ionization range} ($Z_{SIR}$).

This definition remains consistent with the Gaussian beam divergence definition, where at $z_R$, the maximum intensity, $I_{max}(z)$, also decreases by a factor of 2. From a practical standpoint, this approach is advantageous for metrology, as it focuses on estimating the divergence of the ionization profile where the ionization probability is substantial.

Using this new definition for divergence, we find that at $Z_{SIR} = 165.6$ $\mu \text{m}$, where the maximum ionization probability decreases by a factor of 2, the ionization width increases to $w_I(z = 165.6$ $\mu \text{m}) = 4.75$ $\mu \text{m}$. Comparing this to the initial width, $w_I(0) = 4.64$ $\mu \text{m}$, the width has increased by only 2\%. This is a remarkably small divergence compared to the over 40\% (or $\sqrt{2}$) increase in the width of the Gaussian beam.

Such a low divergence has significant implications for strong-field ionization tomography. This implies that while the ionization cross-section $\sigma_i(z)$ varies significantly along the laser beam, the ionization cross-section area $\sigma_\text{SM}(z)$ (or equivalently the radius of ionization, $R_\text{SM}(z)$) remains practically constant.

\begin{figure}[hbpt]
    %\begin{center}
	%\includegraphics[width=\linewidth]{draft_figure_localization}
\includegraphics[width=1\linewidth]{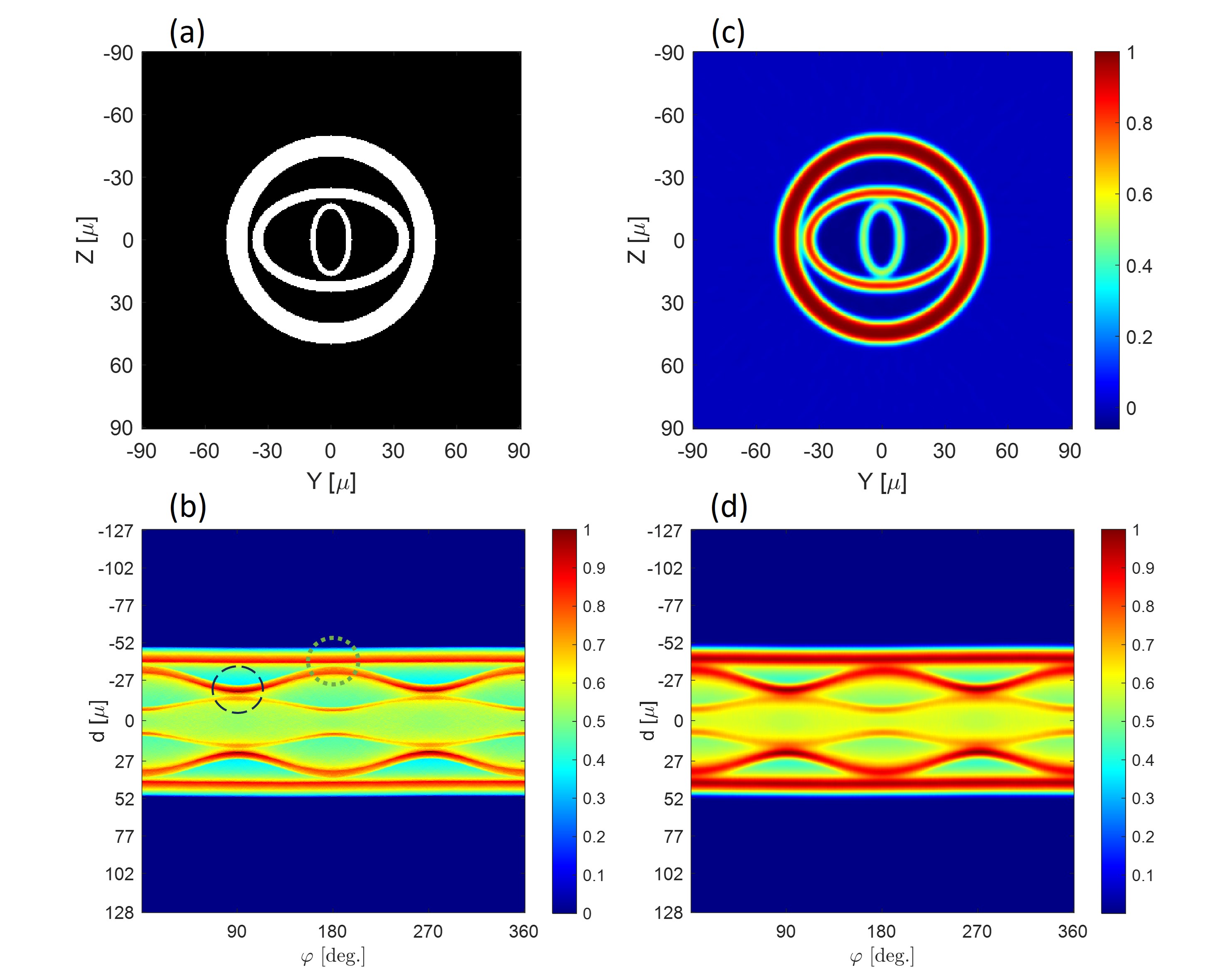}
%"G:\BGU_since_2013\papers\Gas_Jet_ionization_Tomography\Numerical_evaluations_Gas_Jet_ionization\main_tomography_simulation_v2.m“
%The figures are:
%h1r hf5
%H4b h4
% "G:\BGU_since_2013\papers\Gas_Jet_ionization_Tomography\Numerical_evaluations_Gas_Jet_ionization\Figures_making4Theory_Gas_Jet_ionization_paper.pptx"

    %\captionsetup{justification=justified}

    \caption{ \label{Fig_tomography_example} Example of the LISFIT. (a) The target object, defined as a set of three binary objects with uniform density distribution (shown in white) within each object. The assumed target medium ionization potential is $I_p=15.76$  $\textrm{eV}$ (similar to argon).  (b) ``Ideal'' normalized Radon transform of the target object. The green dotted circe and dashed black circle highlight the ``avoided crossing'' due to proxamity of the circular ring (i) to the elliptical ring (ii), and the elliptical ring (ii) to the elliptical ring (iii), respectively.  (c) The  reconstructed object. (d) The normalized total ionization signal as a function of laser beam position, calculated considering relevant laser beam parameters (peak intensity, pulse duration, central wavelength, and beam waist), including diffraction.}

   % \end{center}
\end{figure}	

To explore the implications of our findings, we consider several examples inspired by the well-known Shepp-Logan head phantom \cite{Shepp_FourierHead_IEEE_1974}, commonly used for testing various medical tomography modalities. We begin with a cross-section of a "phantom" object function, defined as a superposition of three simple binary objects, as illustrated in Fig. \ref{Fig_tomography_example} (a): (i) an outer ring with a radius of 50 microns and a width of 10 microns, (ii) an elliptical ring bound by two ellipses, $\left(y/a_{o2}\right)^2+(z/b_{o2})^2=1$ and $\left(y/(a_{o2}-d_2)\right)^2+\left(z/(b_{o2}-d_2)\right)^2=1$, where $a_{o2}=37.5$ $\mu \text{m}$, $b_{o2}=25$ $\mu \text{m}$, and the width $d_{2}=5$ $\mu \text{m}$, and (iii) another elliptical ring similar to (ii), with $a_{o3}=10$ $\mu \text{m}$, $b_{o3}=17.5$ $\mu \text{m}$, and $d_{3}=2.5$ $\mu \text{m}$. 

The calculated Radon transform of this object function is shown in Fig. \ref{Fig_tomography_example} (b). Since the object is symmetrical, the transform exhibits symmetry about the [0, $\varphi$) and [0, d) axes. Three distinct line pairs, with reflection symmetry around the [0, $\varphi$) axis, are clearly identifiable. As the Radon transform is a linear operation, these line pairs correspond to the respective rings that make up the target: the outermost straight line, expected for an object with circular symmetry, corresponds to the transform of the outer ring (i), while the next two curved pairs represent the elliptical rings (ii) and (iii). It is noteworthy that the curves representing rings (i) and (ii) are in close proximity, resembling an "avoided crossing" at angles $0^\circ$, $180^\circ$, and $360^\circ$. This proximity is due to the fact that rings (i) and (ii) are separated by only 2.5 $\mu \text{m}$ along the Y-axis. Similarly, the curves representing elliptical rings (ii) and (iii) are in close proximity at $90^\circ$ and $270^\circ$, as these rings are also separated by just 2.5 $\mu \text{m}$ along the Z-axis.

 In our analysis, we assume a laser beam with a 25 fs FWHM (assuming a Gaussian temporal envelope), centered at $800$ nm, focused at the point (0,0) in the (YZ) plane (Fig. \ref{Fig_tomography_example} (a)). The beam waist is assumed to be $w_0=7$  $\mu \text{m}$,  with a peak intensity of $1.5 \times 10^{14} \text{W}/\text{cm}^2$. We calculate the ionization probabilities Eq. (\ref{prob_ion_after_pulse}) within the laser beam and the total ionization signal Eq. (\ref{Ni_general}) while scanning the angle $\varphi$ and the distance $d$. The results are shown in Fig. \ref{Fig_tomography_example} (a). In this example, the object is well contained within the significant ionization range ($Z_{SIR}=81$ $\mu m$).
 
 While, as expected, the ``ideal'' Radon transform in Fig. \ref{Fig_tomography_example} (b) is similar to the total ionization signal map in Fig. \ref{Fig_tomography_example} (d), there are notable differences. 
 The curves representing each of the three rings in the "exemplary" target become broader and less distinct, particularly at the proximity points  ($0^\circ$, $90^\circ$ , $180^\circ$,  $270^\circ$, and $360^\circ$).
 This broadening and blurring results from convolution with the total ionization probability distribution across the laser beam Eq. (\ref{lamda_Ni}), as previously discussed.

\begin{figure}
    %\begin{center}
    %"G:\BGU_since_2013\papers\Gas_Jet_ionization_Tomography\Numerical_evaluations_Gas_Jet_ionization\main_ionization_profiles_v2.m"
    %h1_xy=figure;
    %"G:\BGU_since_2013\papers\Gas_Jet_ionization_Tomography\Numerical_evaluations_Gas_Jet_ionization\Figures_making4Theory_Gas_Jet_ionization_paper.pptx"
	
\includegraphics[width=0.8\linewidth]{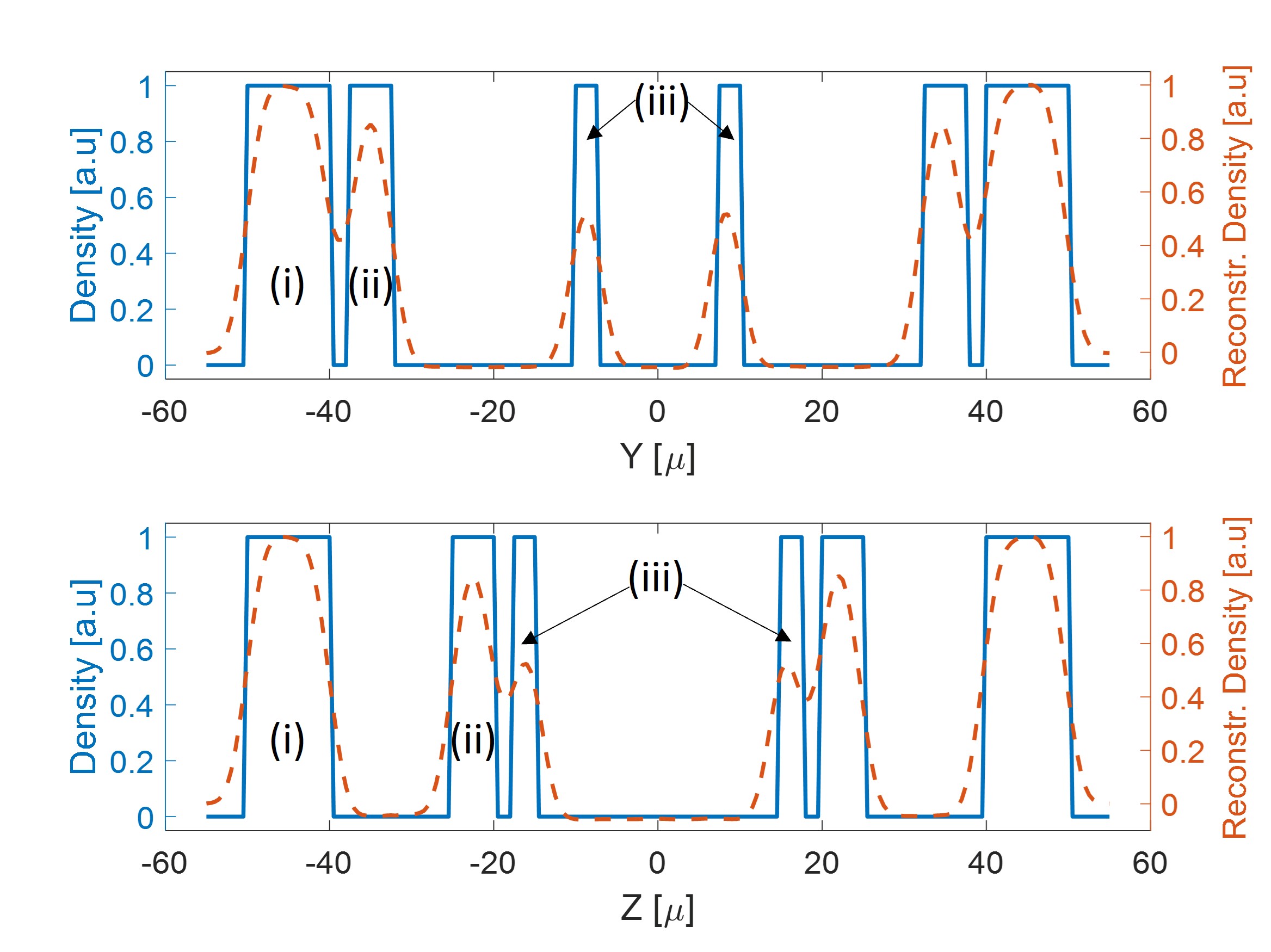}

    \caption{Cross-cuts of the target object (solid blue curve) and the reconstructed object  (dashed red curve) shown in Fig. \ref{Fig_tomography_example}.  The upper panel shows the cross-cut along the (0Y) axis, while the lower panel shows the cross-cut along the (0Z) axis.}

    \label{Fig_tomography_Xsection}
    %\end{center}
\end{figure}

 Using Eqs. \ref{Fourier_lamda_Ni}-\ref{n_0_recostructed}, we reconstruct the target density distrubution, as illustrated in Fig.\ref{Fig_tomography_example} (c). Figure \ref{Fig_tomography_Xsection} depicts cross-cuts of the original and recontructed target along the (0Y) and (0Z) axes.
 This simulation confirms that LISFIT is capable of accurately measuring and reconstructing target objects.
 
 At the same time, the effect of resolution loss due to  finite width $R_{SM}$ Eq. (\ref{r_sigma_2moment}) of the ionization profile across the laser beam is clearly evident. Distinct rings (i) and (ii), separated by only 2.5 $\mu m$ along the Y-axis in Fig. \ref{Fig_tomography_example} (a) (solid blue line, upper panel of Fig. \ref{Fig_tomography_Xsection}), become smeared in the reconstruction (dashed red line in Fig. \ref{Fig_tomography_Xsection}), leading to a loss of contrast. A similar loss of resolution occurs along the (0Z) axis between rings (ii) and (iii) (lower panel of Fig. \ref{Fig_tomography_Xsection}). Additionally, the originally rectangular profiles of the rings become broadened into a bell-like shape, as expected from the convolution of the rectangular function with the bell-like ionization profile Eq. (\ref{lamda_Ni}).

\subsection{Resolution-intensity coupling}
\label{Section_Resolution_Intensity}

It is revealing to consider the ionization width, which defines the resolution of the LISFIT as a function of laser intensity. Let us recall that, in the case of perturbative nonlinear imaging, such as second-order (e.g., SHG \cite{Gannaway_Scanning_Micr_SHG_OQE_1978} or two-photon fluorescence \cite{Denk_TPF_microscopy_Science_1990}) or third-order (e.g., THG \cite{Barad_THG_microscopy_APL1997}) imaging modalities, the resolution is intensity-independent and is given by $w_{2\text{-order}}=w_0/\sqrt{2}$ and $w_{3\text{-order}}=w_0/\sqrt{3}$, respectively. In general, the resolution using an n-order perturbative nonlinear process driven by a laser Gaussian beam is given by $w_{n\text{-order}}=w_0/\sqrt{n}$.

\begin{figure}
    %\begin{center}
    %"G:\BGU_since_2013\papers\Gas_Jet_ionization_Tomography\Numerical_evaluations_Gas_Jet_ionization\main_ionization_profiles_v1.m"
    %"G:\BGU_since_2013\papers\Gas_Jet_ionization_Tomography\Numerical_evaluations_Gas_Jet_ionization\Figures_making4Theory_Gas_Jet_ionization_paper.pptx"
	
\includegraphics[width=0.5\linewidth]{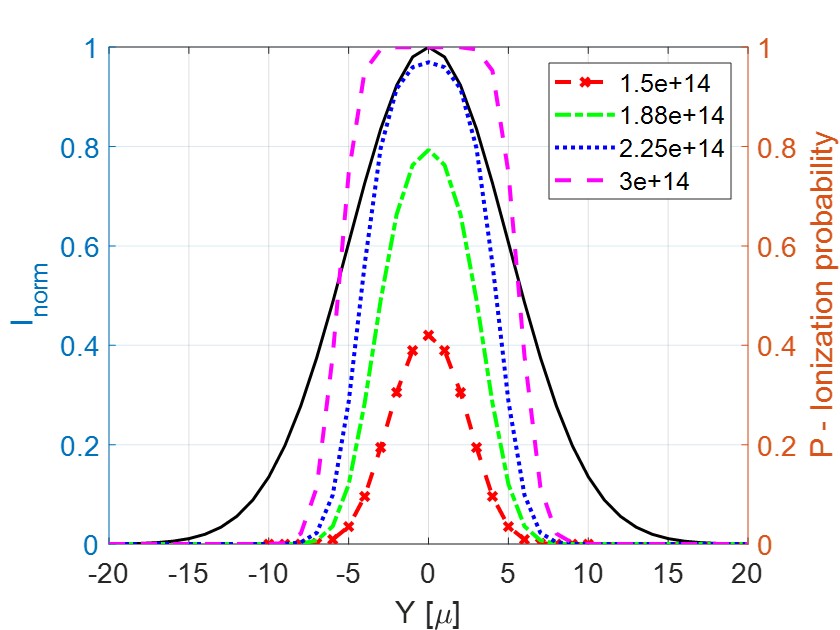}

    \caption{The strong-field ionization probability across the focused laser beam, with a spot size of $w_0=10 \mu m$, as a function of peak intensity.
    Solid black curve represents a normalized Gaussian intensity beam profile across the beam in the focus.
The legend indicates the maximum peak intensity across the Gaussian beam profile in the units of [$W/cm^2$].
Note that as the peak intensity increases, saturation becomes evident in the ionization profile.}

    \label{Fig_ionization_vs_intensity}
    %\end{center}
\end{figure}

Figure \ref{Fig_ionization_vs_intensity} depicts the ionization probability across the laser beam focus as the maximum peak intensity gradually increases, ranging from $I_{peak}=1.5\times 10^{14}$ W/cm\textsuperscript{2} in the far-from-saturation regime to $I_{peak}=3\times 10^{14}$ W/cm\textsuperscript{2}
in the highly saturated regime.
It is noteworthy that the shape of the ionization profile undergoes a transformation, gradually taking a more "squarish" form as the intensity approaches and surpasses the saturation regime.

\begin{figure}
   %"G:\BGU_since_2013\papers\Gas_Jet_ionization_Tomography\Numerical_evaluations_Gas_Jet_ionization\main_ionization_profiles_v1.m"
    %"G:\BGU_since_2013\papers\Gas_Jet_ionization_Tomography\Numerical_evaluations_Gas_Jet_ionization\Figures_making4Theory_Gas_Jet_ionization_paper.pptx"

    %\begin{center}

\includegraphics[width=0.9\linewidth]{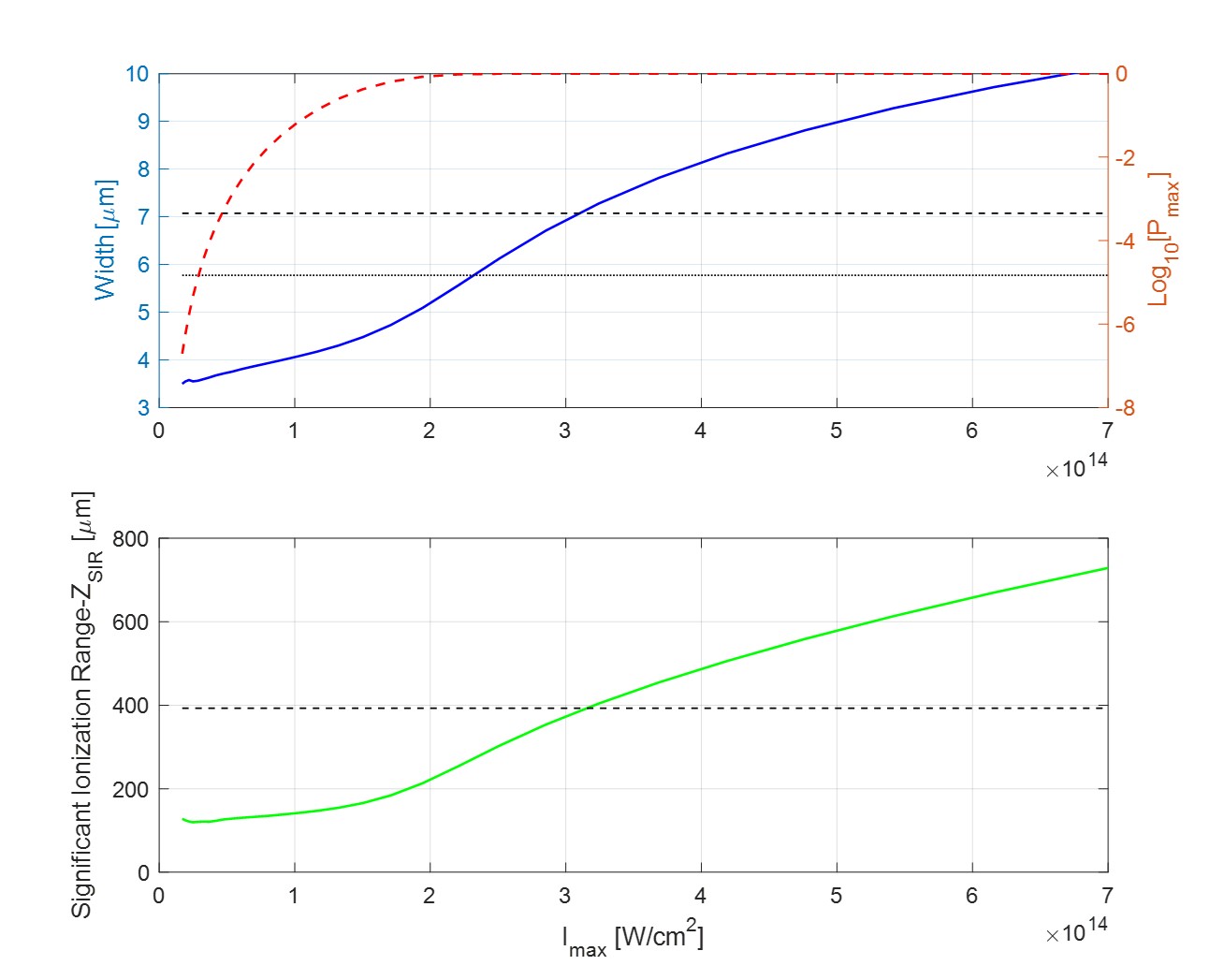}

    \caption{Laser-induced strong field imaging/tomography resolution as a function of intensity. (a) Ionization width, $W_I$  (solid blue line),  shown alongside ionization probability (dashed red curve) versus peak intensity. The horizontal dashed line represents the expected SHG width  ($w_0/\sqrt{2}$), and the dotted line shows the expected THG width ($w_0/\sqrt{3}$). (b) Significant ionization range, $Z_{SIR}$, as a function of peak laser intensity. The horizontal dashed line indicates the Rayleigh range of the driving laser field.}

    \label{Fig_ion_width_vs_peak_intensity}
    %\end{center}
\end{figure}	

In striking contrast to perturbative nonlinear imaging, the resolution of strong-field ionization imaging and tomography is intensity dependent, as evident from Figure \ref{Fig_ion_width_vs_peak_intensity} (a). The width of the ionization profile increases, and the resolution decreases accordingly, as the peak intensity (and probability of ionization) increases.
%[?Discuss SNR vs. resolution. May be put some numbers??]

Obviously, as the intensity increases, the ionization probability and the expected signal also increase. Hence, we have a trade-off between resolution and signal-to-noise ratio (SNR) that we can explicitly and quantitatively assess using the presented approach.

It is worth noting that with careful selection of the driving laser intensity, the expected resolution of SFI can be significantly better than the resolution of both second-order and third-order perturbative imaging modalities, as shown in Fig. \ref{Fig_ion_width_vs_peak_intensity} (a). Specifically, in the discussed emblematic example, choosing the driving laser intensity to be $1.5 \times 10^{14}$ W/cm\textsuperscript{2} results in an SFI width of less than 4.5 $\mu m$, while the resolutions of THG, SHG, and linear imaging are 5.8 $\mu m$, 7 $\mu m$, and 10 $\mu m$, respectively. This SFI resolution is achieved while providing a significant signal with a 0.4 ionization probability at the peak, as shown in Fig. \ref{Fig_ionization_vs_intensity}. This implies that we obtain better than a factor of 10 improvement in the minimum resolvable "voxel" volume for the LISFIT compared to linear tomography modalities (e.g., one-photon fluorescence).

%[?Discuss that Ralley range is also intensity dependent. Implications for imaging and tomography? ]

%??Show example of the plot that have the same $\sigma_0 g(z)$ , but different width of  $P_i(x-x_0,y-y_0,z)$ (may be use Gaussian driving beam and show the $P_i(x-x_0,y-y_0,z)$ at the focus and out of focus with higher intensity.

%??Relate these finding to tomography resolution??

%?? Relate resolution to intensity and to ionization rate. Plot ionization probability as function of the driving intensity (for example for the 25fsec laser beam)??.
%Make the SNR discussion??

%?? Show the difference between different models in the context of ionization width and ionization yield applying Gaussian beam as the driver. Compare to Gaussian beam.
%Future studies:Make generalization for Gaussian beams with $\textrm{M}^2 \neq 1$. Probably follow Siegman approach for definition of Gaussian beam with $\textrm{M}^2 \neq 1$. ??

\subsection{Localization and apodization}
\label{Section_Localization}

Another characteristic feature stemming from the nonlinearity of strong-field interactions is the phenomenon of localization. Up to this point, we have considered the case where the extent of the measured target in the (YZ) plane is much smaller than the Rayleigh range of the driving laser beam. Based on our discussion of ionization profile properties, we can refine this criterion by replacing the Rayleigh range of the laser beam with the significant ionization range, $Z_{SIR}$. In regions significantly larger than $Z_{SIR}$, ionization — and, consequently, the contribution to the total ionization signal in Eq. (\ref{Ni_general}) — becomes negligible.

This gives rise to an inherent localization property in strong-field ionization tomography. Such localization can be highly useful, for example, when measuring a gas jet target in the presence of a buffer gas. In this scenario, the background signal contribution from the buffer gas is inherently suppressed, a feature unique to the LISFIT. This contrasts sharply with linear tomographic modalities, where background gas can degrade the SNR and even introduce artifacts in the tomographic reconstruction.

Figure \ref{Fig_ion_width_vs_peak_intensity} (b) shows the dependence of $Z_{SIR}$ on the driving laser intensity. Notably, decreasing the intensity simultaneously improves both resolution and localization (i.e., reduces $Z_{SIR}$), while also decreasing the ionization signal. Thus, our analysis provides an essential toolbox for optimizing key parameters: spatial resolution, localization, and SNR.

 This localization property is somewhat analogous to the optical sectioning found in laser scanning perturbative nonlinear microscopy \cite{Denk_TPF_microscopy_Science_1990}. In both cases, the nonlinearity causes a rapid reduction in the generated signal for out-of-focus beams, and the total signal generated across the laser beam (which corresponds to the ionization cross-section, $\sigma_i(z)$, in our analysis) decreases, as shown in Fig. \ref{Fig_ionization_profile_v1} (a).

However, there are important differences. In microscopy, focusing is typically done with a much higher numerical aperture (NA), and the optical sectioning achieved by nonlinearity often enhances longitudinal resolution to the point that it serves as an alternative to confocal microscopy \cite{Denk_TPF_microscopy_Science_1990}. 
On the other hand, in the LISFIT, we typically employ much looser focusing, resulting in a substantially longer $Z_{SIR}$, which localizes the extent of the measured target along the laser beam propagation direction.

\begin{figure}[hbpt]
    %\begin{center}
	%\includegraphics[width=\linewidth]{draft_figure_localization}
\includegraphics[width=0.95\linewidth]{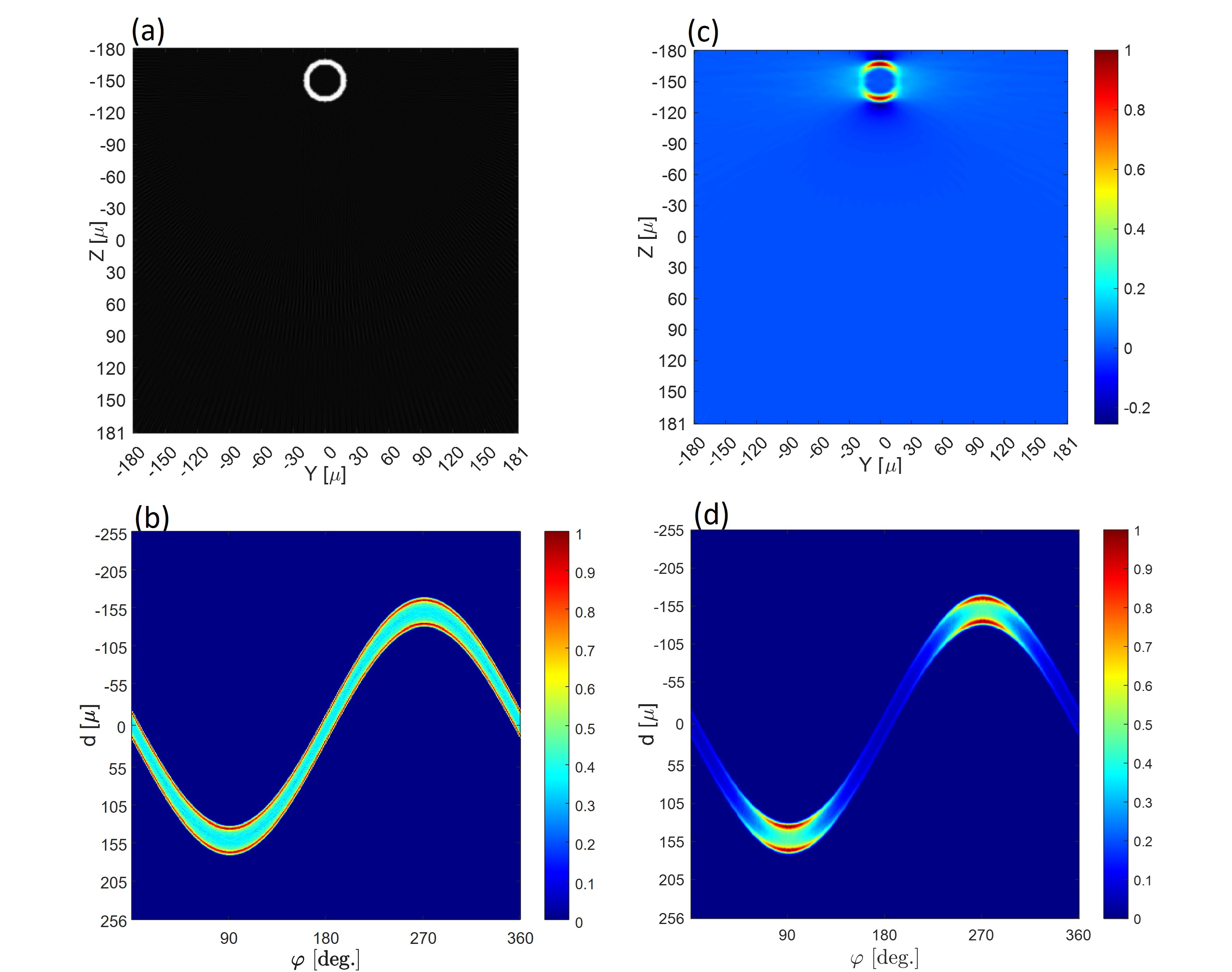}
%"G:\BGU_since_2013\papers\Gas_Jet_ionization_Tomography\Numerical_evaluations_Gas_Jet_ionization\main_tomography_simulation_v1.m“
%From: main_tomography_simulation_v1.m the figures  copied and edited in edited in "G:\BGU_since_2013\papers\Gas_Jet_ionization_Tomography\Numerical_evaluations_Gas_Jet_ionization\Figures_making4Theory_Gas_Jet_ionization_paper.pptx"

    %\captionsetup{justification=justified}

    \caption{ \label{Fig_Tomography_Localization_effect} (a) Tomographic reconstruction of the object cross section with an "ideal" non-diffracting uniform ionization profile.  (b) Tomographic parallel projection map for an "ideal" non-diffracting uniform ionization profile. (c) Tomographic reconstruction of the target object (ionization potential of Ar, $I_p = 15.76 eV$, is assumed) with a strong-field ionization profile driven by a laser beam with a central wavelength $\lambda_0 = 800 nm$,  waist focused at (x,z)=(0,0), $w_0 = 7 \mu m$, pulse duration $\tau_{FWHM}=25 fsec$ with a Gaussian temporal profile with maximum intensity of $1.5 X 10^{14}$ W/cm\textsuperscript{2}. (d) Tomographic parallel projection map for a a strong-field ionization profile as described in (c).}

   % \end{center}
\end{figure}

 To explore what happens when the extent of the measured target in the (YZ) plane extends beyond $Z_{SIR}$, we consider an object function in the form of a small ring with a uniform density distribution. The ring has a diameter of 20 $\mu m$ and width of 5 $\mu m$, shifted by 150 $\mu m$ from the center of the focused laser beam, as illustrated in Fig. \ref{Fig_Tomography_Localization_effect} (a). 
 The laser, with the same parameters and target $I_p$ as in the example shown in Fig. \ref{Fig_tomography_example} (where $Z_{SIR}=81$ $\mu m$), is focused at (0, 0) along the (YZ) plane. Tomographic scanning is performed by rotating the target around the X-axis ((Y,Z) = (0,0)), and for each angle, the scanning is achieved by moving the laser beam across the Y-axis.
 
 The tomographic parallel projection map of the normalized Radon transform (equivalent to a measurement with an "ideal" non-diffracting beam) is shown in Fig. \ref{Fig_Tomography_Localization_effect} (b). As expected, this map exhibits a uniform signal distribution.
 
Figure \ref{Fig_Tomography_Localization_effect} (d) illustrates the normalized total ionization signal as a function of the laser beam position, denoted by $d$, and the scanning direction $\varphi$,  using the relevant laser beam parameters similar to the measurements shown in Fig. \ref{Fig_tomography_example} (d). The scanning range $d$ across the laser beam has been chosen to exceed the extent of the target from the origin in the (Y,Z) plane.

It is evident that in the latter case, the parallel projections signal is highly non-uniform as a function of $\varphi$. For $\varphi\simeq 90^\circ$ and 
 $270^\circ$, the measured parallel projection signal has clear maxima. This is because, at these angles, the scanning laser beam crosses the target near its waist, well within the $Z_{SIR}$ range, where the ionization cross-section $\sigma_i(z)$, is close to its maximum.
 Conversely,  for $\varphi\simeq 0^\circ$ and $180^\circ$, the laser beam crosses the target at the greatest distance from the waist ($\thicksim$150 $\mu m$), beyond the $Z_{SIR}$ range, where the  $\sigma_i(z)$ is minimal within the scanning area. As a result, the measured parallel projection signal exhibits clear minima at these angles. This is a manifistation of the apodization phenomenon.
 
 It is important to note that, while the parallel projection signal exhibits strong non-uniformity as a function of the scanning angle $\varphi$, the overall shape of each parallel projection varies relatively little. This can be attributed to the previously discussed observation that, for strong-field interaction driven by a Gaussian laser beam, the ionization cross-sectional area $\sigma_\text{SM}(z)$ remains almost constant along the laser beam, while the ionization cross-section $\sigma_i(z)$ varies significantly (see Fig. \ref{Fig_ionization_profile_v1}).
 
 Figure \ref{Fig_Tomography_Localization_effect} (c) shows the tomographic reconstruction of the object using the strong-field ionization parallel projection map shown in Fig. \ref{Fig_Tomography_Localization_effect} (d). As a result of the apodization phenomenon, the reconstructed target displays strong non-uniformity and shape distortion near the minima, even though the original target was fully uniform.

%Localization phenomena in the ([?"theta-d" plane. Describe in details with simulation for demonstration. Only for specific angles (90 degrees and 90+180=270 degrees far distant objects full projection dynamic range is preserved  ?]).
%Apodization, assymptotically it is like partial angle data tomographic available.
%??Compare with sectioning in SHG and THG?? - in the next paper

\section{Conclusion}

In summary, this study presents a comprehensive theoretical analysis of the  LISFIT. We identified a unique coupling between resolution and intensity, demonstrating LISFIT's potential to achieve superior resolution compared to traditional linear and perturbative nonlinear imaging techniques. Our findings highlight the advantages of strong-field tomography, particularly its ability to reduce background noise, a capability rooted in the localization effect inherent to the nonlinearity of strong-field ionization physics.

The newly defined significant ionization range $Z_{SIR}$ establishes a crucial criterion for resolving localized ionization  with minimal divergence, underscoring the robustness of LISFIT in mulidimensional imaging applications.
 Additionally, we identified and demonstrated the effect of apodization, which may cause reconstruction distortions if the localization region is not properly chosen during experimental design.
 
These findings provide a foundation for optimizing experimental design by balancing key parameters such as resolution, signal strength, signal-to-noise ratio, and localization extent.
Future experimental research may build upon these findings to expand the practical applications of LISFIT in diverse contexts. By leveraging insights gained from this study, LISFIT is positioned to become a powerful tool for high-resolution multidimensional imaging in various scientific and technological applications.

Future studies may also consider investigating the effects of double ionization, potential laser beam propagation distortions in high-density media, and developing advanced reconstruction algorithms to mitigate the impact of apodization.
%
%??? Mention in the future:
%
%** It is important to note, however, that under certain conditions,
%double ionization may play a role [28], which will be a subject for future studies.
%
%**For the sake of clarity and simplicity we will assume undepleted  pump and neglect beam distortions within the measured medium in our analysis.   [?Note in conclusions or somewhere that we can do without these assumptions at the expense of mathematical complexity?] 
%
%**[? Discuss the fact that ionization profile $P_i(x,y,z)$ changes along propagation axis. We will first assume that it is not changing significantly, making the approximation \ref{Ni_general}. Mention that the more detailed analysis of z-varient profile of $P_i(x,y,z)$ is a topic for future studies in the spirit of tomographic imaging with diffracting sources. Discuss an apodization and localization ?] - largely done alreay.
%Double ionization

Acknowledgements: E.F. acknowledges the support of Israel Science Foundation (ISF) grant 2855/21.

\section{Appendix}

\subsection{Appendix: Strong field ionization}
\label{Appendix_SFI}

Here we summarise ionization rate formulas used throughout the paper.

Ionization rate $W_{PPT}(E_L, \omega_L)$ in the Perelomov, Popov, Terentev (PPT) \cite{Perelomov_PPT_JETP1966} model:
\begin{multline}
\label{PPT_ionization}
W_{PPT}(E_L, \omega_L) = C_{n^* l^*}^2 f_{lm} I_p \left( \frac{2}{E_L n^{*3}} \right)^{2n^*-|m|-1} \\
\times (1+ \gamma^2)^{|m|/2+3/4} A_m (\omega_L, \gamma) \\
\times \exp{\left( - \frac{2(2 I_p )^{3/2}}{3 E_L} g (\gamma)  \right)},
\end{multline}

where:

$n^*$  effective principal quantum number:
\begin{equation}
\label{n_star_factor}
 n^* = \frac{Z}{\sqrt{2 I_p}},
 \end{equation}

$l^*$ effective orbital quantum number:
\begin{equation}
\label{l_star_factor}
 l^* =n^* - 1 ,
 \end{equation}

Good discussion and summary for the choice of effective quantum numbers can be found in \cite{Ilkov_Ionization_atoms_tunnel_J_Phys_B_1992}.

\begin{equation}
\label{c_2_factor}
 C_{n^* l^*}^2 = \frac{2^{2n^*}}{n^* \Gamma (n^*+l^*+1) \Gamma (n^* -l^*)},
 \end{equation}

 \begin{equation}
\label{f_lm_factor}
 f_{lm} = \frac{(2l+1)(l+|m|)!}{2^{|m|} |m|! (l-|m|)!},
 \end{equation}

\begin{equation}
\label{A_factor}
A_m (\omega_L, \gamma)= \frac{4 }{\sqrt{3 \pi}|m|!} \frac{\gamma ^2}{1+ \gamma ^2} \sum_{k >= k_{min}}^{+\infty}  e^{-\alpha (\gamma) (k- k_{min})} \xi _m \left( \sqrt{\beta (\gamma) (k- k_{min})} \right)
\end{equation}

\begin{equation}
\label{alpha_factor}
 \alpha (\gamma) =  2 \left[ \textrm{arcsinh} ( \gamma) - \frac{\gamma}{\sqrt{1 + \gamma ^2}} \right] ,
 \end{equation}

\begin{equation}
\label{beta_factor}
 \beta (\gamma) = \frac{2 \gamma}{\sqrt{1 + \gamma ^2}} ,
 \end{equation}

\begin{equation}
\label{xi_factor}
 \xi _m (x) = \frac{x ^{2|m|+1}}{2} \int^1_0 \frac{e^{-x^2 t} t^{|m|}}{\sqrt{1-t}} \, dt ,
 \end{equation}

\begin{equation}
\label{g_factor}
 g(\gamma) = \frac{3}{2 \gamma} \left[\left(1+ \frac{1}{2 \gamma ^2} \right) \textrm{arcsinh} ( \gamma) - \frac{\sqrt{1 + \gamma ^2}}{2 \gamma} \right],
 \end{equation}

where k is the number of Above threshold ionization photons (ATI), starting with the minimum number of ATI photons, $k_min$ (this formula for clearity is written in the SI units):
\begin{equation}
\label{k_min_photons}
k_{min} = \lceil{\frac{I_p + U_p}{\hbar \omega_L}}\rceil=\lceil{\frac{I_p}{\hbar \omega}\left({1+\frac{1}{2 \gamma^2}}\right)}\rceil; U_p= \frac{e^2 E_L^2}{4 m_{\textrm{e}} \omega_L^2},
\end{equation}

Here, $U_p$ is a ponderomotive energy, which is the cycle averaged kinetic energy of the electron motion in the field.

The ADK formula for the ionization rate $W_{ADK}(E_L, \omega_L)$ :

\begin{equation}
\label{ADK_ionization_rate}
W_{ADK}(E_L, \omega_L) = C_{n^* l^*}^2 f_{lm} I_p \left( \frac{2}{E_L n^{*3}} \right)^{2n^*-|m|-1}
\times \exp{\left( - \frac{2(2 I_p )^{3/2}}{3 E_L}  \right)},
\end{equation}

\subsection{Appendix: Ionization probability}

\label{Appendix_probability}

Let's express the total probability of ionization  $P_i(t)$ during time window  $(-\infty,t]$, via ionization rate $W(t)$. Let's $\overline{P}_i(t)$ to  denote the complement event to $P_i(t)$ , i.e. the probability of no ionization occurring in the time window $(-\infty,t]$. Then, we have:

\begin{equation}
\label{prob_ion_full}
P_i(t)+\overline{P_i}(t)=1 \Rightarrow  \partial{P_i (t)} /\partial{t} = - \partial{\overline{P}_i(t)}/\partial{t}
\end{equation}

Following basic  theorem for the joint probability, we can write:

\begin{equation}
\label{prob_ion2}
 P_i(t+\triangle t)=P_i(t)+\overline{P_i}(t) W(t) \triangle t \Rightarrow \partial{P_i(t)}/\partial{t}  = \overline{P}_i(t) W(t)
\end{equation}

Substituting (\ref{prob_ion_full}) into (\ref{prob_ion2}), we obtain:

\begin{equation}
\label{prob_ion3}
  \partial{\overline{P_i}(t)}/\partial{t}  = -\overline{P}(t) W(t) \Rightarrow \frac{d \overline{P}(t)}{\overline{P}(t)} = - W(t) dt
\end{equation}

Integrating (\ref{prob_ion3}) and taking into account that $\overline{P_i}(-\infty)=1$ (probability of not being ionized before the laser pulse arrived is equal to 1), we get:

\begin{equation}
\label{prob_ion4}
   \int_{-\infty}^{t} \frac{d \overline{P_i}(t')}{\overline{P_i}(t')} = - \int_{-\infty}^{t} W(t') dt' \Rightarrow \ln \overline{P_i}(t) = - \int_{-\infty}^{t} W(t') dt' \Rightarrow \overline{P_i}(t) = \exp \left[ {- \int_{-\infty}^{t} W(t') dt'} \right]
\end{equation}

 %$\overline{p}(t)$ - probability that no ionization occured in the time period, hence $\overline{p}(t) +p(t)=1$
%
%
%$ \partial{p(t)}/\partial{t} \triangle t + o(\triangle t^2) =\overline{p}(t) w(t) = (1-p(t)) w(t) $
%$ \partial{p(t)}/\partial{t}  = \overline{p}(t) w(t) = (1-p(t)) w(t) $

Finally, plugging (\ref{prob_ion4})  into (\ref{prob_ion_full}),  obtain:

\begin{equation}
\label{prob_ion5}
P_i(\overrightarrow{\textbf{r}},t) = 1 - \exp \left[ {- \int_{-\infty}^{t} W(\overrightarrow{\textbf{r}},t') dt'} \right]
\end{equation}

\bibliography{G:/BGU_since_2013/papers/Atto_references_since_2024_10_11}
%\bibliography{C:/Users/Eugene/BGU_since_2013/papers/Atto_references_since_2024_10_11}

\end{document}